\documentclass[pre,aps,preprint,amsmath]{revtex4}
\usepackage{fancyhdr}
\usepackage{graphicx}
\usepackage{dcolumn}
\usepackage{txfonts}
\usepackage{bm}
\begin{document}

\title{Diffusion metamaterials for plasma transport}

\author{Fubao Yang}
\email{18110190009@fudan.edu.cn}
\affiliation{Department of Physics, State Key Laboratory of Surface Physics, and Key Laboratory of Micro and Nano Photonic Structures (MOE), Fudan University, Shanghai 200438, China}
\author{Zeren Zhang}
\affiliation{Department of Physics, State Key Laboratory of Surface Physics, and Key Laboratory of Micro and Nano Photonic Structures (MOE), Fudan University, Shanghai 200438, China}

\begin{abstract}
 Plasma technology has found widespread applications in numerous domains, yet the techniques to manipulate plasma transport predominantly rely on magnetic control. In this review, we present a streamlined diffusion-migration method to characterize plasma transport. Based on this framework, the viability of the transformation theory for plasma transport is demonstrated. Highlighted within are three model devices designed to cloak, concentrate, and rotate plasmas without significantly altering the density profile of background plasmas. Additionally, insights regarding potential implications for novel physics are discussed. This review aims to contribute to advancements in plasma technology, especially in sectors like medicine and chemistry.
\end{abstract}

\maketitle
\section{Introduction} 

Since ancient times, energy has always been one of the core needs for human development. Over the past few centuries, people have developed and used various sources of energy, such as fossil fuels, hydropower, and nuclear energy. However, with the growth of the world's population and economy, the demand for energy has also been continuously increasing, posing significant challenges to global energy resources. Now, we face a deepening energy crisis, including the depletion of fossil resources and the severe environmental impacts of carbon emissions. Against this backdrop, the sustainability of energy and how to efficiently use existing energy resources have become hot topics of global concern, urgently requiring solutions. Therefore, energy management and regulation have become crucial. We need to explore various measures to achieve active control of energy, thereby using it more efficiently. In this process, the development of metamaterials has become a prominent field and is one of the key means to achieve active energy control.

Metamaterials refer to novel materials that achieve extraordinary properties through artificial structures. Their origin can be traced back to the phenomenon of negative refraction proposed by Professor Veselago in 1968~\cite{Yfb-Ve}. However, due to the limited material and technological levels at that time, this phenomenon went unverified experimentally for a long time. It wasn't until the end of the 20th century that Professor Pendry's team demonstrated the feasibility of negative refraction experimentally through metal wire arrays and split-ring resonator structures~\cite{Yfb-PendryPRL96,Yfb-PendryPRL99}. Since then, the research on electromagnetic metamaterials has attracted widespread attention. Particularly, the establishment of transformation optics theory in 2006 provided a powerful tool for designing electromagnetic metamaterial devices with novel functionalities, enabling the free manipulation of electromagnetic wave propagation as desired~\cite{Yfb-LeoSci06,Yfb-PenSci06}. This marks a new height in mankind's active control over physical fields, greatly propelling the development of the electromagnetic metamaterials field~\cite{Yfb-ChenNM10,Yfb-ZheNM12,Yfb-KadicRPP13,Yfb-FuPQE19}.

The successful application of metamaterials in wave systems laid a solid foundation for their expansion into diffusion systems. In 2008, Professor Huang Jiping's team from Fudan University established the transformation thermotics in the heat conduction system and predicted the functionality of a thermal cloak~\cite{Yfb-FanAPL2008}, inaugurating the field of diffusion metamaterials~\cite{Yfb-ZhangNRP23,Yfb-rmp}. Over the next decade, transformation theory continually developed in diffusion systems~\cite{Yfb-LiuEPJAP2009, Yfb-LiuCTP10, Yfb-QiuAIP Adv.15, Yfb-Tan15, Yfb-ShenPRL16, Yfb-HuangPB17, Yfb-JiIJMPB18, Yfb-XuEPJB18, Yfb-XuJAP18, Yfb-XuPLA18, Yfb-HuangPP18, Yfb-HuangESEE19, Yfb-Huang20, Yfb-XuPRAP20, Yfb-XuESEE20, Yfb-HuangESEE20, Yfb-XuIJHMT20, Yfb-HuangPhysics20, Yfb-WangiScience20, Yfb-YangPR21, Yfb-LeiEPL21, Yfb-ZhangATS22, Yfb-WangCPB22, Yfb-YangPRAP22, Yfb-XuBook23, Yfb-YangPRA23, Yfb-DaiPR23,Yfb-Tan20}
, with an increasing number of extended theories proposed, including effective medium theory~\cite{Yfb-FanCTP10, Yfb-ZhaoFOP12, Yfb-ChenEPJAP15, Yfb-ShenAPL16, Yfb-YangAPL17, Yfb-XuEPJB17, Yfb-WangJAP17, Yfb-ShangIJHMT18, Yfb-JiCTP18, Yfb-DaiEPJB18, Yfb-ShangJHT18, Yfb-WangIJTS18, Yfb-XuPRE18, Yfb-WangPRA19, Yfb-XuPRA19a, Yfb-YangPRE19, Yfb-ZhouESEE19, Yfb-DaiIJHMT20, Yfb-WangPRE20, Yfb-XuEPJB20, Yfb-WangPRAP20, Yfb-XuAPL20, Yfb-XuCPLEL20, Yfb-WangICHMT20, Yfb-DaiJNU21, Yfb-XuEPL21, Yfb-XuPRE21, Yfb-ZhangTSEP21, Yfb-TianIJHMT21, Yfb-XuAPL21, Yfb-XuEPL21-2, Yfb-HuangAMT22, Yfb-XuPRL22, Yfb-LinSCPMA22, Yfb-ZhuangPRE22, Yfb-XuPRL22-2, Yfb-ZhouEPL23, Yfb-LeiIJHMT23, Yfb-ZhuangIJMSD23, Yfb-XuNSR23, Yfb-LeiMTP23, Yfb-XuPANS23,Yfb-ZhangCPL23,Yfb-click,Yfb-research,Yfb-nc}
, scattering cancellation theory~\cite{Yfb-XuEPJB19, Yfb-XuEPL19, Yfb-XuPRA19, Yfb-YangEPL19, Yfb-XuEPJB19a, Yfb-YangEPL19a, Yfb-XuPRE19, Yfb-XuSCPMA20, Yfb-SuEPL20, Yfb-QuEPL21, Yfb-JinPNAS23, Yfb-ZhangPRA23}
, and optimization algorithms~\cite{Yfb-LiuJAP21,Yfb-LiuJAP21-2,Yfb-ZhangPRD22,Yfb-am}, among others. Accordingly, various diffusion metamaterial devices were designed or fabricated, such as thermal cloak~\cite{Yfb-ChenAPL2008,Yfb-GaoJAP2009,Yfb-LiJAP10,Yfb-SuFOP11,Yfb-QiuEPL13,Yfb-QiuIJHT14,Yfb-Tan16,Yfb-YangJAP19,Yfb-XuCPL20,Yfb-WangATE21,Yfb-WangPRAP21,Yfb-XuIJHMT21,Yfb-XuPRAP19,Yfb-XuEPL20a,Yfb-YangJAP20,Yfb-XuPRE20,Yfb-DaiPRAP22,Yfb-YaoISci22}
, concentrator~\cite{Yfb-Tan4,Yfb-Tan11,Yfb-DaiPRE18,Yfb-WangJAP18,Yfb-YangESEE19,Yfb-ZhuangSCPMA22}
, sensor~\cite{Yfb-JinIJHMT20,Yfb-XuEPL20b,Yfb-JinIJHMT21,Yfb-WangEPL21}
, or rotator~\cite{Yfb-YangPRAP20,Yfb-QiuCTP14,Yfb-DaiJAP18}. Diffusion metamaterials have provided a novel approach for controlling the diffusion of energy and matter and have become a current hotspot in research. The in-depth study of diffusion metamaterials may also bring inspiration and vitality to classical thermodynamic and statistical physics~\cite{Yfb-HuangCP04,Yfb-HuangPRE04d,Yfb-HuangPRE04b,Yfb-HuangPRE05a,Yfb-XuPLA06,Yfb-ChenJPA07,Yfb-WangPNAS09,Yfb-WangOL10,Yfb-ZhaoPNAS11,Yfb-SongARCS12,Yfb-liang2013pre,Yfb-Liang2013fopw,Yfb-QiuFP14,Yfb-QiuPLA14,Yfb-QiuCPB14,Yfb-Tan12,Yfb-Tan5,Yfb-Tan2,Yfb-XinPA17,Yfb-XinFP17,Yfb-JiPA18} or soft condensed matter physics~\cite{Yfb-HuangJMMM05, Yfb-HuangSSC00, Yfb-HuangCTP01-2, Yfb-HuangCTP01-1, Yfb-PanPB01, Yfb-HuangPRE01, Yfb-HuangJPCM02, Yfb-HuangPRE02, Yfb-HuangCTP02, Yfb-HuangPLA02, Yfb-HuangCTP03, Yfb-HuangPRE03-2, Yfb-GaoPRE03, Yfb-HuangJAP03, Yfb-HuangPRE03-1, Yfb-GaoEPJB03,Yfb-DongJAP04,Yfb-KoJPCM04, Yfb-GaoPRB04, Yfb-HuangPRE04g, Yfb-LiuPLA04, Yfb-HuangCPL04, Yfb-HuangPRE04f, Yfb-HuangJPCM04, Yfb-HuangPLA04, Yfb-KoEPJE04, Yfb-HuangAPL04, Yfb-HuangPRE04e, Yfb-HuangJCP04, Yfb-DongJAP04-2, Yfb-HuangPRE04c, Yfb-HuangJPCB04, Yfb-WangJAP03, Yfb-ShenCPL06, Yfb-WangCPL06, Yfb-HuangAPL06, Yfb-HuangJPD06, Yfb-WangPRE06, Yfb-ZhangPRE06, Yfb-ZhangCPL07, Yfb-ZhangPRE07}.
\begin{figure}[!ht]
	\centering
	\includegraphics[width=.9\linewidth]{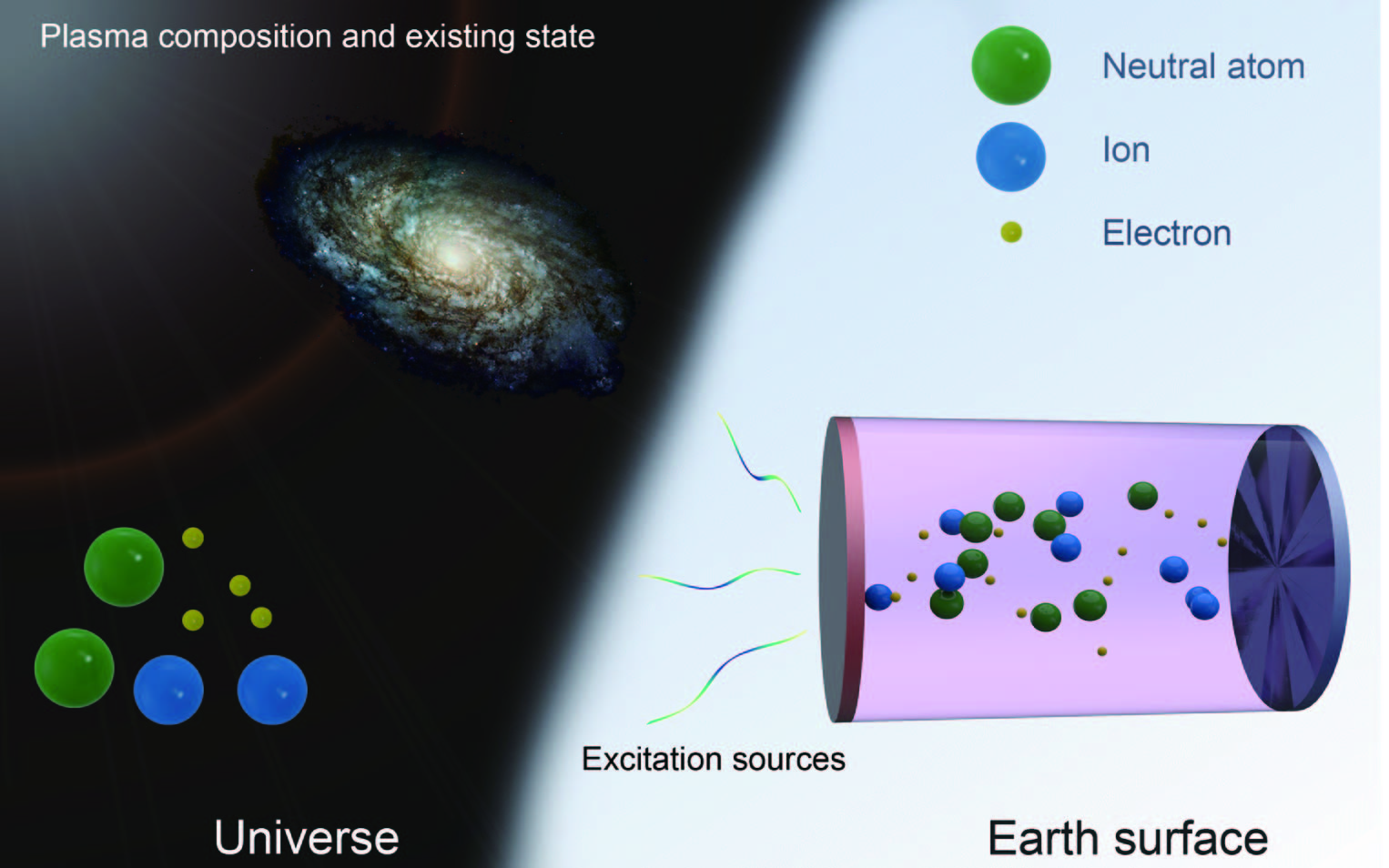}
	\caption{Schematic diagrams of plasma components and distribution.}
	\label{yfb-fig1}
\end{figure}

Plasma diffusion, as a specific form of mass diffusion, has seen limited progress in the research of metamaterials designed for its regulation. Plasma, often referred to as the fourth phase of matter, consists of a gaseous blend of free ions, electrons, and various reactive species, granting it strong electrical conductivity. While it's not typically found on Earth's crust, plasma can be synthetically produced by electrifying gases using direct/alternating currents or sources of radio/microwave frequencies. See Fig.~\ref{yfb-fig1}. Given its distinct nature, plasma-based technologies hold significant importance across sectors including micro/nanoelectronics, chemical research, bio-medicine, aerospace, and materials science~\cite{Yfb-SamalJCP17,Yfb-LiangAEM18,Yfb-TamIEEE20,Yfb-LiAST21}. While there has been a plethora of theoretical and empirical research, mastering the control over plasma transportation remains a formidable task. Traditional methods of directing charged particles primarily rely on external magnetic fields, a straightforward approach that may curtail precise control. Over recent years, the transformation theory, which substitutes spatial transformation with material alteration, has gained prominence in wave and diffusion systems as an effective strategy to steer matter or energy trajectories~\cite{Yfb-HuAM18,Yfb-HuAM19,Yfb-ZhangCPL21}. Yet, its application in the domain of plasma transportation, which can be viewed as a distinct diffusion mechanism, remains unexplored. The intricate motion dynamics within plasmas might explain this gap. Only recently, reports have suggested the potential application of transformation theory to plasma transport~\cite{Yfb-ZhangCPL22}. Although this work merely used a toy model to describe the plasma diffusion process, it has paved the way for new possible approaches to control plasma.
\section{Transformation Theory for Plasma Transport}
\subsection{For steady-state plasma transport}
Compared to traditional diffusion systems, the transport in actual plasmas is more complex. This is because collisions between charged particles, and between charged and neutral particles, often result in reactions. Furthermore, the intrinsic electric field within the plasma can affect the movement of charged particles. Here, we do not consider collisions that result in nuclear reactions. Hence, the governing equation for the transport of charged particles in plasma can be written as:
\begin{subequations}\label{c4md1}
	\begin{align}
		\partial_{t} n+& \mathbf{\nabla}\cdot\Gamma =0,\\
		\mathbf{\nabla}\cdot\Gamma =-\mathbf{\nabla}\cdot\left(D\cdot \mathbf{\nabla} n\right)& \pm \mathbf{\nabla}\cdot \left(\mu \boldsymbol{E}n\right)+\mathbf{\nabla}\cdot \left(\boldsymbol{v}n\right),\label{c4md1-1}
	\end{align}
\end{subequations}
where $n$, $D$, $\mu$, $\boldsymbol{E}$, and $\boldsymbol{v}$ represent the plasma density, diffusion coefficient, mobility, electric field, and advective velocity, respectively. The first term on the right-hand side of equation (\ref{c4md1-1}) represents the diffusion term. The second term represents the mobility term, with the positive sign indicating the mobility of cations and the negative sign indicating the mobility of anions or electrons. The third term denotes the effect of convection. The transport equation for plasma also needs to be coupled with the Poisson's equation:
\begin{equation}\label{c4poss}
	\boldsymbol{\nabla}\cdot\boldsymbol{E}=\frac{e}{\epsilon_0}\left(n_i-n_e \right),
\end{equation}
where $e$, $\epsilon_0$, $n_i$, and $n_e$ represent the elementary charge, permittivity of free space, cation density, and electron density, respectively. Assuming the plasma is electrically neutral internally, the right-hand side of the above equation is zero. For simplicity, we will not consider the advective process for now~\cite{Yfb-CuiJAP19}. Therefore, equation (\ref{c4md1}) can be written as:
\begin{equation}\label{c4md2}
	\partial_{t} n-\mathbf{\nabla}\cdot\left(D\cdot \mathbf{\nabla} n\right) \pm \mathbf{\nabla}\cdot \left(\mu \boldsymbol{E}n\right)=0.
\end{equation}
Note that in plasmas, the relationship between the diffusion coefficient and mobility is given by the Einstein relation:
\begin{equation}\label{c4ein}
	\mu=\frac{De}{Tk_B}=\frac{D}{T_i},
\end{equation}
where $T$ is the particle temperature, $k_B$ is the Boltzmann constant, and $T_i=Tk_B/e$ is the reduced temperature in volts (V). Assuming the temperature is constant, the reduced temperature is also constant. Using the Einstein relation, we can simplify parameters and further reduce the plasma transport equation to:
\begin{equation}\label{c4md3}
	\partial_{t} n-\mathbf{\nabla}\cdot\left(D \cdot\mathbf{\nabla} n\right) \pm \mathbf{\nabla}\cdot \left[\left(\frac{D\cdot \boldsymbol{E}}{T_i}\right)n\right]=0.
\end{equation}
Next, we will study the transformation theory in plasmas based on the above equation. First, we discuss the transport under steady-state conditions.

In the steady state, where the plasma density does not vary with time, equation~(\ref{c4md3}) becomes:
\begin{equation}\label{c4mds1}
	-\mathbf{\nabla}\cdot\left(D\cdot \mathbf{\nabla} n\right) \pm \mathbf{\nabla}\cdot \left[\left(\frac{D\cdot \boldsymbol{E}}{T_i}\right)n\right]=0.
\end{equation}
To visualize the transformation results more intuitively, the equation above is converted into its component form in curved space~\cite{Yfb-DaiFP21}:
\begin{equation}\label{c4mds2}
	-\partial_i \left(\sqrt{A}D^{ij} \partial_j n\right)\pm \partial_i \left[\left(\frac{\sqrt{A}D^{ij} E_j}{T_i}\right)n\right]=0.
\end{equation}
The curved space corresponds to coordinates $x_i$, with its covariant base vectors denoted as $\boldsymbol{a}_i$ and $\boldsymbol{a}_j$. In equation~(\ref{c4mds2}), $A$ is the determinant of $\boldsymbol{a}_i\cdot \boldsymbol{a}_j$. The transformed equation~(\ref{c4mds2}) in physical space is expressed as:
\begin{equation}\label{c4mds3}
	-\partial_{i'}\left[\frac{\partial x'_i}{\partial x_i}\sqrt{A}D^{ij} \frac{\partial x'_j}{\partial x_j}\partial_{j'} n\mp\frac{\partial x'_i}{\partial x_i} \left(\frac{\sqrt{A}D^{ij} E_j}{T_i}\right)n\right]=0,
\end{equation}
Here, the physical space coordinates are denoted as $x'_i$. Thus, $\partial x'_i/\partial x_i$ and $\partial x'_j/\partial x_j$ are components of the Jacobian matrix $\mathbf{J}$, and the determinant of this Jacobian matrix satisfies $\det \mathbf{J}=1/\sqrt{A}$. This equation can then be rewritten as:
\begin{equation}\label{c4mds4}
	-\boldsymbol{\nabla}'\cdot\left(\frac{\mathbf{J}D\mathbf{J}^\tau}{\det\mathbf{J}}\cdot \boldsymbol{\nabla}' n\right)\pm \boldsymbol{\nabla}'\cdot \left(\frac{\mathbf{J}D\cdot \boldsymbol{E}}{\det\mathbf{J}T_i}n\right)=0,
\end{equation}
We can simplify equation~(\ref{c4mds4}) by incorporating the extra metric through physical parameters, resulting in:
\begin{subequations}\label{c4mds5}
	\begin{align}
		-\boldsymbol{\nabla}'\cdot\left(D'\cdot \boldsymbol{\nabla}' n\right)\pm &\boldsymbol{\nabla}'\cdot \left(\frac{D'\cdot \boldsymbol{E}'}{T_i}n\right)=0,\label{c4mds5-1}\\
		D'&=\frac{\mathbf{J}D\mathbf{J}^\tau}{\det \mathbf{J}},\label{c4mds5-2}\\
		\boldsymbol{E}'&=\mathbf{J}^{-\tau}\boldsymbol{E}.\label{c4mds5-3}
	\end{align}
\end{subequations}
Here, $\boldsymbol{\nabla}'$ indicates the differential in the new coordinate system; $\mathbf{J}^{\tau}$ is the transpose of $\mathbf{J}$; $\mathbf{J}^{-\tau}$ represents the inverse of $\mathbf{J}^{\tau}$. Comparing equation~(\ref{c4mds1}) and (\ref{c4mds5-1}) shows they have the same form, implying that the migration diffusion equation in the steady state strictly maintains its transformation form. Hence, we can design parameter control for steady-state plasma transport using equations~(\ref{c4mds5-2}) and (\ref{c4mds5-3}).

\subsection{For transient-state plasma transport }
Although the transformation theory fits well with steady-state plasma transport, the situation under transient conditions is quite different. Re-examining equation~(\ref{c4md3}) and processing it in a manner similar to the steady-state, we obtain:
\begin{equation}\label{c4mdt1}
	\frac{1}{\det\mathbf{J}}\partial_t n-\boldsymbol{\nabla}'\cdot\left(\frac{\mathbf{J}D\mathbf{J}^\tau}{\det\mathbf{J}}\cdot \boldsymbol{\nabla}' n\right)\pm \boldsymbol{\nabla}'\cdot \left(\frac{\mathbf{J}D\cdot \boldsymbol{E}}{\det\mathbf{J}T_i}n\right)=0,
\end{equation}
Substituting equations~(\ref{c4mds5-2}) and (\ref{c4mds5-3}) into equation~(\ref{c4mdt1}), we obtain:
\begin{equation}\label{c4mdt2}
	\frac{1}{\det\mathbf{J}}\partial_t n-\boldsymbol{\nabla}'\cdot\left(D'\cdot \boldsymbol{\nabla}' n\right)\pm \boldsymbol{\nabla}'\cdot \left(\frac{D'\cdot \boldsymbol{E}'}{T_i}n\right)=0,
\end{equation}
Comparing equations~(\ref{c4md3}) and (\ref{c4mdt2}), we notice an irreducible parameter in front of the time-dependent term. The transformed equation has changed its form, indicating that the transformation theory fails for transient plasma transport. Approximating equation~(\ref{c4mdt2}), we get:
\begin{subequations}\label{c4mdt3}
	\begin{align}
		\partial_t n-\boldsymbol{\nabla}'\cdot\left(D''\cdot \boldsymbol{\nabla}' n\right)&\pm \boldsymbol{\nabla}'\cdot \left(\frac{D''\cdot \boldsymbol{E}''}{T_i}n\right)=0,\label{c4mdt3-1}\\
		D''&=\mathbf{J}D\mathbf{J}^{\tau},\label{c4mdt3-2}\\
		\boldsymbol{E}''&=\mathbf{J}^{-\tau}\boldsymbol{E}.\label{c4mdt3-3}
	\end{align}
\end{subequations}
Where equations~(\ref{c4mdt3-2}) and (\ref{c4mdt3-3}) are parameter transformation rules. In this manner, the equation retains its transformation-invariant characteristics. Similar to the convection-diffusion equation, equation~(\ref{c4mdt3-1}) is an approximate form of equation~(\ref{c4md3}). Only when $\det\mathbf{J}=1$, does equation~(\ref{c4mdt3-1}) strictly transform to equation~(\ref{c4md3}).

Revisiting equation~(\ref{c4mdt1}) and multiplying both sides by $\det\mathbf{J}$, we get:
\begin{equation}\label{c4mdt4}
	\partial_t n -\det\mathbf{J}\left[\boldsymbol{\nabla}'\cdot\left(\frac{\mathbf{J}D\mathbf{J}^\tau}{\det\mathbf{J}}\cdot \boldsymbol{\nabla}' n\mp \frac{\mathbf{J}D \cdot\boldsymbol{E}}{\det\mathbf{J}T_i}n\right)\right]=0,
\end{equation}
Then, decomposing the differential terms and isolating $1/\det\mathbf{J}$, we get:
\begin{equation}\label{c4mdt5}
	\begin{split}
		&\det\mathbf{J}\left[
		\mathbf{\nabla}'\left(\frac{1}{\det\mathbf{J}}\right)\cdot\left(\mathbf{J}D\mathbf{J}^\tau\cdot\mathbf{\nabla}'n\right)+
		\frac{1}{\det\mathbf{J}}\mathbf{\nabla}'\cdot\left(\mathbf{J}D\mathbf{J}^\tau\cdot\mathbf{\nabla}'n\right)\right] \\
		&\mp\det\mathbf{J}\left[\mathbf{\nabla}'\left(\frac{1}{\det\mathbf{J}}\right)\cdot\left(\frac{\mathbf{J}D\cdot\boldsymbol{E}}{T_i}n\right)+
		\frac{1}{\det\mathbf{J}}\mathbf{\nabla}'\cdot\left(\frac{\mathbf{J}D\cdot\boldsymbol{E}}{T_i}n\right)\right]=\partial_t n
	\end{split}
\end{equation}
Combining the terms that include $\mathbf{\nabla}'\left(1/\det \mathbf{J} \right)$, we get:
\begin{subequations}\label{c4mdt6}
	\begin{align}
		\partial_t n &=\mathbf{\nabla}'\cdot\left(\mathbf{J}D\mathbf{J}^\tau\cdot\mathbf{\nabla}'n\mp\frac{\mathbf{J}D\cdot\boldsymbol{E}}{T_i}n\right)+\Delta,\label{c4mdt6-1}\\
		\Delta &=\det\mathbf{J}\mathbf{\nabla}'\left(\frac{1}{\det\mathbf{J}} \right)\cdot\left(\mathbf{J}D\mathbf{J}^{\tau}\cdot \mathbf{\nabla}'n\mp \frac{\mathbf{J}D\cdot\boldsymbol{E}}{T_i} n \right).\label{c4mdt6-2}
	\end{align}
\end{subequations}
We observe that if $\Delta$ is sufficiently small to be ignored, then equation~(\ref{c4mdt6-1}) reverts to equation~(\ref{c4mdt3-1}). Now, analyzing the specific expression of the error in equation~(\ref{c4mdt6-2}), which involves the Jacobian determinant, plasma diffusion coefficient, plasma density, temperature, and electric field. The spatial transformation determines the form of the Jacobian. Hence, $\Delta$ is related to the physical parameters of the system and the specific form of spatial transformation. To minimize the error, both diffusion coefficients and electric field need to be sufficiently small. Moreover, there exist special spatial transformations for which $\det\mathbf{J}=1$, making the error $\Delta=0$. In this case, equation~(\ref{c4mdt3-1}) is strictly the same as equation~(\ref{c4md3}).

\section{Potential Applications for Transformation-based plasma metamaterials}
\begin{figure}[!ht]
	\centering
	\includegraphics[width=.9\linewidth]{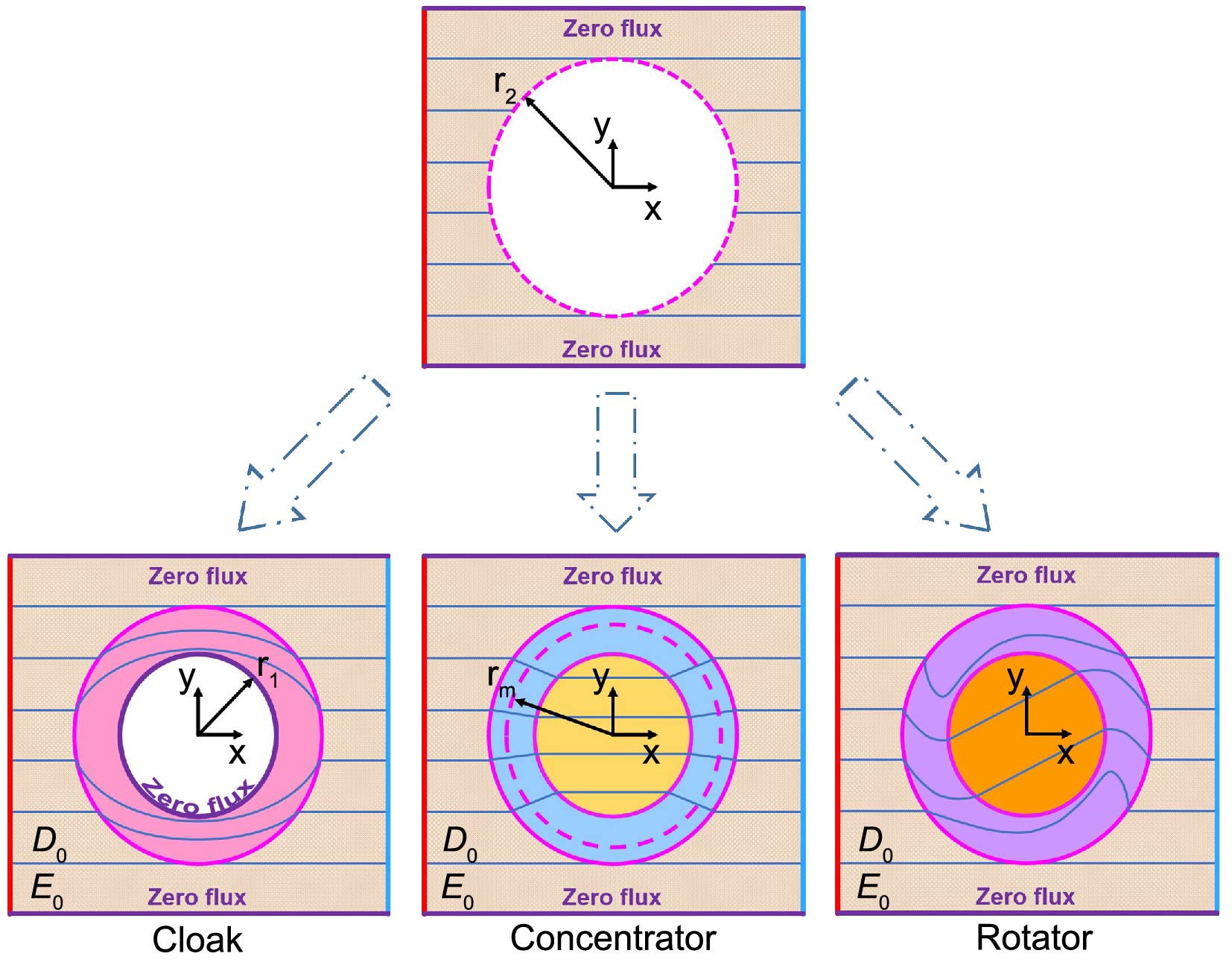}
	\caption{Schematic diagrams of the three plasma devices. From left to right: cloak, concentrator, and rotator. The blue solid lines represent the direction of the plasma flow. The side length of the square background in the model is set to $l=0.12$~m. The diffusivity of the background is set as $D_0=9.2\times10^{-7}$~m~$s^{-1}$, and the field strength is $E_{x0}=1.04\times10^4$~V~m$^{-1}$. Other parameter values: $r_1=0.020$~m, $r_2=0.030$~m, $r_m=0.025$~m, $\theta_0=\pi/3$, and the reduced temperature is set to $T_0=2.0$~V. Adapted from Ref.~\cite{Yfb-ZhangCPL22}.}
	\label{c4fig2}
\end{figure}

To validate our theory, we introduce three conceptual plasma devices: the plasma cloak, concentrator, and rotator. See Figure~\ref{c4fig2}. A unique feature they share is that while achieving their respective functionalities, they do not affect the plasma distribution in the background medium, which is quite distinctive in the plasma field. Specifically, the cloak can protect the core region, meaning that the background does not affect the core, and vice versa; the concentrator can concentrate plasma flow, enhancing the density gradient in the core region; the rotator can alter the propagation direction of the plasma in the core region. Next, we will detail the coordinate transformation relations for implementing the functionalities of these three devices, starting with the cloak.

\subsection{Cloak}
To realize the plasma cloak, we can express the coordinate transformation relationship from the virtual space $r_i$ to the physical space $r'_i$ as
\begin{equation}\label{c4ct1}
	\begin{aligned}
		r' &=\frac{r_2-r_1}{r_2}r+r_1,\\
		\theta' &=\theta,
	\end{aligned}
\end{equation}
where $r_1$ and $r_2$ represent the inner and outer diameters of the cloak, respectively. Identical to the transformation relationship of the chemical wave cloak in the previous work, the spatial transformation here also expands the point at the center in the virtual space into an inner circle, thereby compressing the outer circle into a ring with an inner diameter equivalent to the radius of the small circle. We can then compute the Jacobian matrix based on equation~(\ref{c4ct1}), expressed as
\begin{equation}\label{c4ja1}
	\textbf{J}=\left[
	\begin{matrix}
		\frac{r_2-r_1}{r_2} & 0\\
		0 & \frac{\left( r_2-r_1\right) r'}{r_2\left(r'-r_1 \right)}
	\end{matrix}
	\right].
\end{equation}
Next, based on equations~(\ref{c4mdt3-2}) and (\ref{c4mdt3-3}), the transformation parameters required for the cloak can be calculated as
\begin{subequations}\label{c4pa1}
	\begin{align}
		D''&=\left[
		\begin{array}{cc}
			D_0\left( \frac{r_2-r_1}{r_2}\right) ^2 & 0\\
			0 & D_0\left[\frac{\left( r_2-r_1\right) r'}{r_2\left(r'-r_1 \right)}\right]^2
		\end{array}
		\right], \\
		\boldsymbol{E}''&=\left[
		\begin{array}{cc}
			\frac{r_2}{r_2-r_1}E_r \\
			\frac{r_2\left(r'-r_1 \right)}{\left( r_2-r_1\right) r'}E_t
		\end{array}
		\right],
	\end{align}
\end{subequations}
where $D_0$ is the diffusion rate of the background, and $E_r$ and $E_t$ are the radial and tangential components of the background electric field $\boldsymbol{E}_0$, respectively. In this model, the background electric field is set in the $x$ direction, so $\boldsymbol{E}_0=\left[E_{x0},\,0\right]^\tau$~V~m$^{-1}$. Thus, based on the conversion relationship between Cartesian coordinates and cylindrical coordinates, the specific forms of $E_r$ and $E_t$ are
\begin{equation}\label{c4elec}
	\boldsymbol{E}_0=\left[
	\begin{array}{cc}
		E_r \\
		E_t
	\end{array}
	\right]=\left[
	\begin{array}{cc}
		E_{x0}\cos\theta\\
		-E_{x0}\sin\theta
	\end{array}
	\right].
\end{equation}
Since the transient mass transfer theory is an approximation, the electric field is not continuous at the boundary between the cloak and the background medium. One potential solution is to artificially control the potential at this boundary. Then, with equation~(\ref{c4pa1}), we can achieve transient plasma cloaking.

\subsection{Concentrator}

For the concentrator, the spatial transformation corresponding to the coordinate transformation can be designed as,
\begin{equation}\label{c4ct2}
	\begin{aligned}
		r' &=\frac{r_1}{r_m}r,\qquad r<r_m\\
		r' &=\frac{r_1-r_m}{r_2-r_m}r_2+\frac{r_2-r_1}{r_2-r_m}r,\quad r_m<r<r_2\\
		\theta' &=\theta.
	\end{aligned}
\end{equation}
$r_m$ is a constant between $r_1$ and $r_2$ (represented by a dashed line in figure~\ref{c4fig2}). Similarly, this spatial transformation can be understood in virtual space, where a circle with radius $r_2$ is divided into two parts by a circle with radius $r_m$. One part is the inner circle with a radius of $r_m$ and the remaining part is an annulus with an inner diameter of $r_m$ and an outer diameter of $r_2$. The radius of the inner circle is then compressed to $r_1$ and the inner diameter of the annulus is stretched to $r_1$.

For convenience, let $p=\frac{r_2-r_1}{r_2-r_m}$, $q=\frac{r_1-r_m}{r_2-r_m}r_2$, and $f=\frac{r_1}{r_m}$. Based on equation~(\ref{c4ct2}), the Jacobian matrix of the concentrator is obtained as,
\begin{subequations}\label{c4ja2}
	\begin{align}
		\textbf{J}_1 &=\left[
		\begin{matrix}
			f & 0\\
			0 & f
		\end{matrix}
		\right],~~~~~~~r'<r_1\\
		\textbf{J}_2 &=\left[
		\begin{matrix}
			p & 0\\
			0 & \frac{r'p}{r'-q}
		\end{matrix}
		\right],~~r_1<r'<r_2
	\end{align}
\end{subequations}
According to equations~(\ref{c4mdt3-2}) and~(\ref{c4mdt3-3}), the transformation parameters of the concentrator at $r'<r_1$ ($D_1''$ and $\boldsymbol{E}_1''$) and $r_1<r'<r_2$ ($D_2''$ and $\boldsymbol{E}_2''$) are obtained as,
\begin{subequations}\label{c4pa2}
	\begin{align}
		D_1''& =\left[
		\begin{array}{cc}
			D_0f^2 & 0\\
			0 & D_0f^2
		\end{array}
		\right],\\
		\boldsymbol{E}_1''&=\left[
		\begin{array}{cc}
			E_r/f\\
			E_t/f
		\end{array}
		\right],\\
		D_2''&=\left[
		\begin{array}{cc}
			D_0p^2 & 0\\
			0 & D_0\left(r'p/\left(r'-q\right)\right)^2
		\end{array}
		\right],\\
		\boldsymbol{E}_2''&=\left[
		\begin{array}{cc}
			E_r/p \\
			E_t\left(r'-q\right)/r'p
		\end{array}
		\right].
	\end{align}
\end{subequations}
The efficiency of the concentrator is related to the value of $r_m/r_1$. The larger the $r_m$, the higher the concentration efficiency, corresponding to a larger density gradient. This can be understood from the physical image of the space transformation. When $r_m$ is larger, the compressed area of the inner circle is larger, which corresponds to a higher degree of concentration. With equation~(\ref{c4pa2}), we can achieve the transient plasma concentration.

\subsection{Rotator}

For the rotator, the spatial transformation corresponding to the coordinate transformation is given by,
\begin{equation}\label{c4ct3}
	\begin{aligned}
		r' &=r,\\
		\theta' &=\theta+\theta_0,~~~r<r_1\\
		\theta' &=\theta+\theta_0\frac{r-r_2}{r_1-r_2},~~~r_1<r<r_2
	\end{aligned}
\end{equation}
where $\theta_0$ is a constant rotation angle. This spatial transformation can be understood as a series of circles with radius $r\in[r_1,\, r_2]$ rotating around their center, with their rotation angles linearly varying with their sizes.

For convenience, let $g=\theta_0/\left(r_1-r_2\right)$. The Jacobian matrix of the rotator can then be obtained as,
\begin{subequations}\label{c4ja3}
	\begin{align}
		\textbf{J}_1 &=\left[
		\begin{matrix}
			1 & 0\\
			0 & 1
		\end{matrix}
		\right],~~~~~~~r'<r_1\\
		\textbf{J}_2 &=\left[
		\begin{matrix}
			1 & 0\\
			r'g & 1
		\end{matrix}
		\right].~~~r_1<r'<r_2
	\end{align}
\end{subequations}
Noting that $\det\mathbf{J}_1=\det\mathbf{J}_2=1$, the equation~(\ref{c4mdt3}) becomes an exact solution. Similarly, using equations~(\ref{c4mdt3-2}) and (\ref{c4mdt3-3}), the transformation parameters for the rotator in the regions $r'<r_1$ ($D_1''$ and $\boldsymbol{E}_1''$) and $r_1<r'<r_2$ ($D_2''$ and $\boldsymbol{E}_2''$) can be computed as,
\begin{subequations}\label{c4pa3}
	\begin{align}
		D_1''& =\left[
		\begin{array}{cc}
			D_0 & 0\\
			0 & D_0
		\end{array}
		\right],\\
		\boldsymbol{E}_1''&=\left[
		\begin{array}{cc}
			E_r\\
			E_t
		\end{array}
		\right],\\
		D_2''&=\left[
		\begin{array}{cc}
			D_0 & D_0r'g\\
			D_0r'g & D_0\left[\left(r'g\right)^2+1\right]
		\end{array}
		\right],\\
		\boldsymbol{E}_2''&=\left[
		\begin{array}{cc}
			E_r-r'gE_t\\
			E_t
		\end{array}
		\right].
	\end{align}
\end{subequations}
In practical simulations, the $\boldsymbol{E}''$ must be rotated. Since the parameters after transformation are taken from the new coordinate system, and the new coordinates are rotated by a certain angle relative to the old coordinates, the equation~(\ref{c4pa3}) cannot be used directly if the system's coordinate system is not adjusted. The core parameters in the old cylindrical coordinate system should be,
\begin{equation}\label{c4pa3-1}
	D_1'' =\left[
	\begin{array}{cc}
		D_0 & 0\\
		0 & D_0
	\end{array}
	\right],\quad \boldsymbol{E}_1''=\left[
	\begin{array}{cc}
		E_r\left( \theta-\theta_0\right) \\
		E_t\left( \theta-\theta_0\right)
	\end{array}
	\right],
\end{equation}
and the shell parameters are,
\begin{equation}\label{c4pa3-2}
	D_2''=\left[
	\begin{array}{cc}
		D_0 & D_0rg\\
		D_0rg & D_0\left[\left(rg\right)^2+1\right]
	\end{array}
	\right], \quad \boldsymbol{E}_2''=\left[
	\begin{array}{cc}
		E_r\left( \theta-\theta_r\right)-rgE_t\left( \theta-\theta_r\right)\\
		E_t\left( \theta-\theta_r\right)
	\end{array}
	\right],
\end{equation}
where $E_r\left( \theta-\theta_0\right)$ and $E_t\left( \theta-\theta_0\right)$ represent the values of $E_r$ and $E_t$ at $\theta=\theta-\theta_0$; $E_r\left( \theta-\theta_r\right)$ and $E_t\left( \theta-\theta_r\right)$ represent the values of $E_r$ and $E_t$ at $\theta=\theta-\theta_r$; and $\theta_r=g\left(r-r_2\right)$. $E_r$ and $E_\theta$ are determined by equation~(\ref{c4elec}). The transformation form of the diffusion rate is not affected by the rotation of the coordinate system, because rotation of the scalar $D$ has no meaning. Up to this point, we have derived the parameter transformation rules for regulating the plasma, namely the cloak, the concentrator, and the rotator, which are (\ref{c4pa1}), (\ref{c4pa2}), (\ref{c4pa3-1}) and (\ref{c4pa3-2}). Finite-element simulation is used to verify the model.

\subsection{Simulation Verification}
\begin{figure}[!ht]
	\centering
	\includegraphics[width=.9\linewidth]{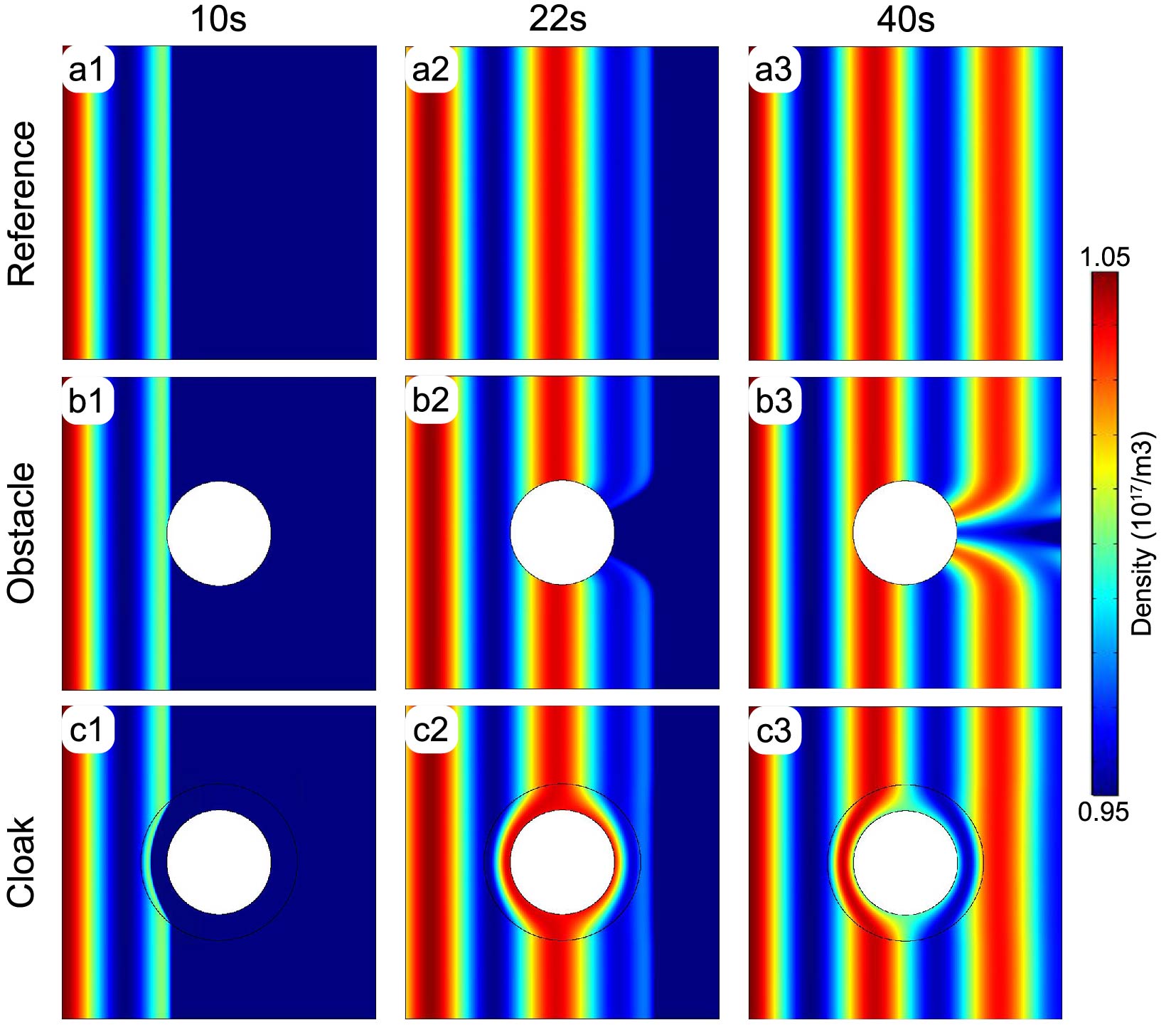}
	\caption{Simulation results of the cloak at transient states. (a1)-(a3) Density profiles for pure background at 10 s, 22 s, and 40 s, respectively. (b1)-(b3) Density profiles for background with an obstacle at 10 s, 22 s, and 40 s, respectively. (c1)-(c3) Density profiles for background with the cloak at 10 s, 22 s, and 40 s, respectively. Adapted from Ref.~\cite{Yfb-ZhangCPL22}.}
	\label{yfb-fig2}
\end{figure}

\begin{figure}[!ht]
	\centering
	\includegraphics[width=.9\linewidth]{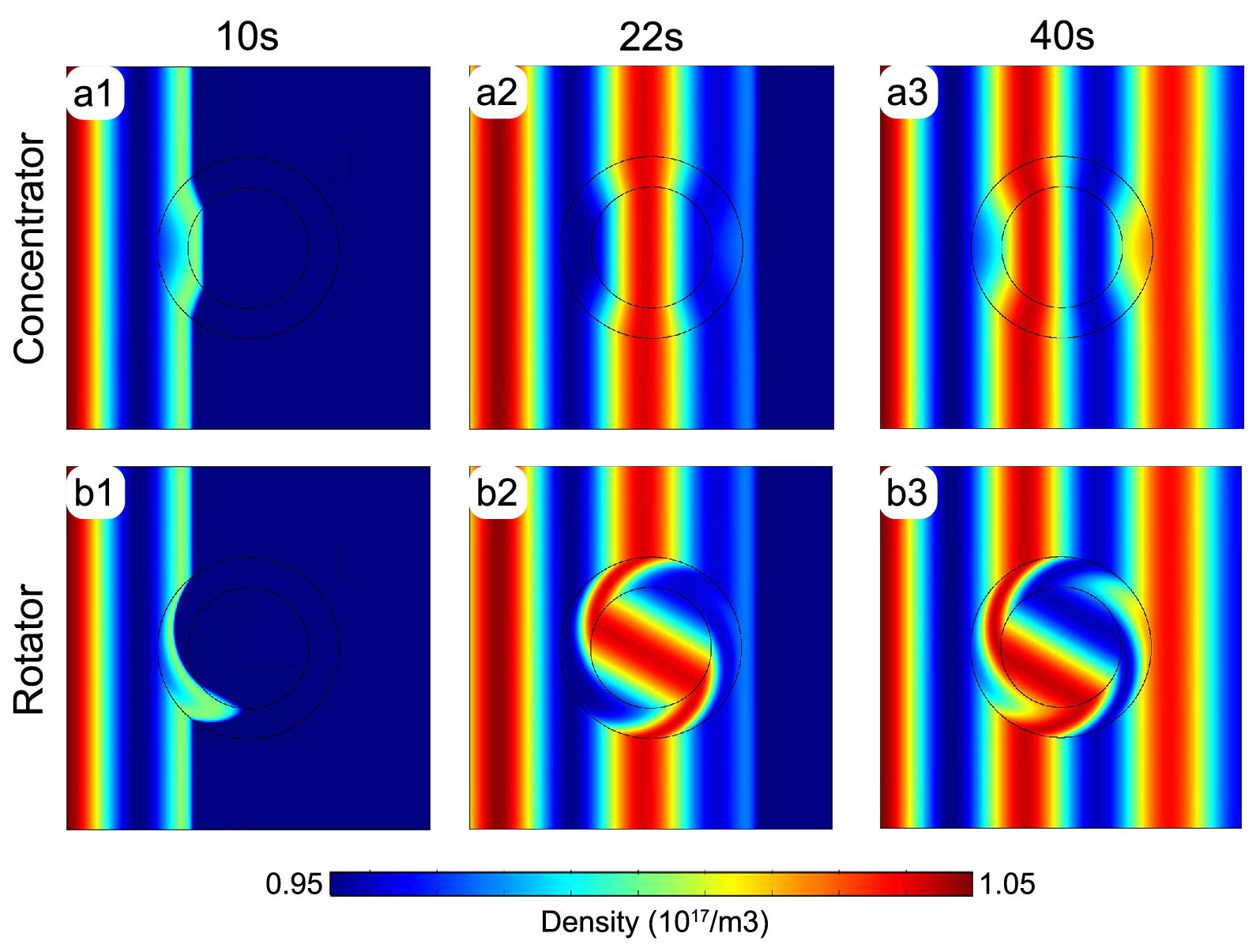}
	\caption{Simulation results of concentrator and rotator at transient states. (a1)-(a3) Density profiles for the concentrator at 10 s, 22 s, and 40 s, respectively. (b1)-(b3) Density profiles for the rotator at 10 s, 22 s, and 40 s, respectively. Adapted from Ref.~\cite{Yfb-ZhangCPL22}.}
	\label{yfb-fig3}
\end{figure}

\begin{figure}[!ht]
	\centering
	\includegraphics[width=.9\linewidth]{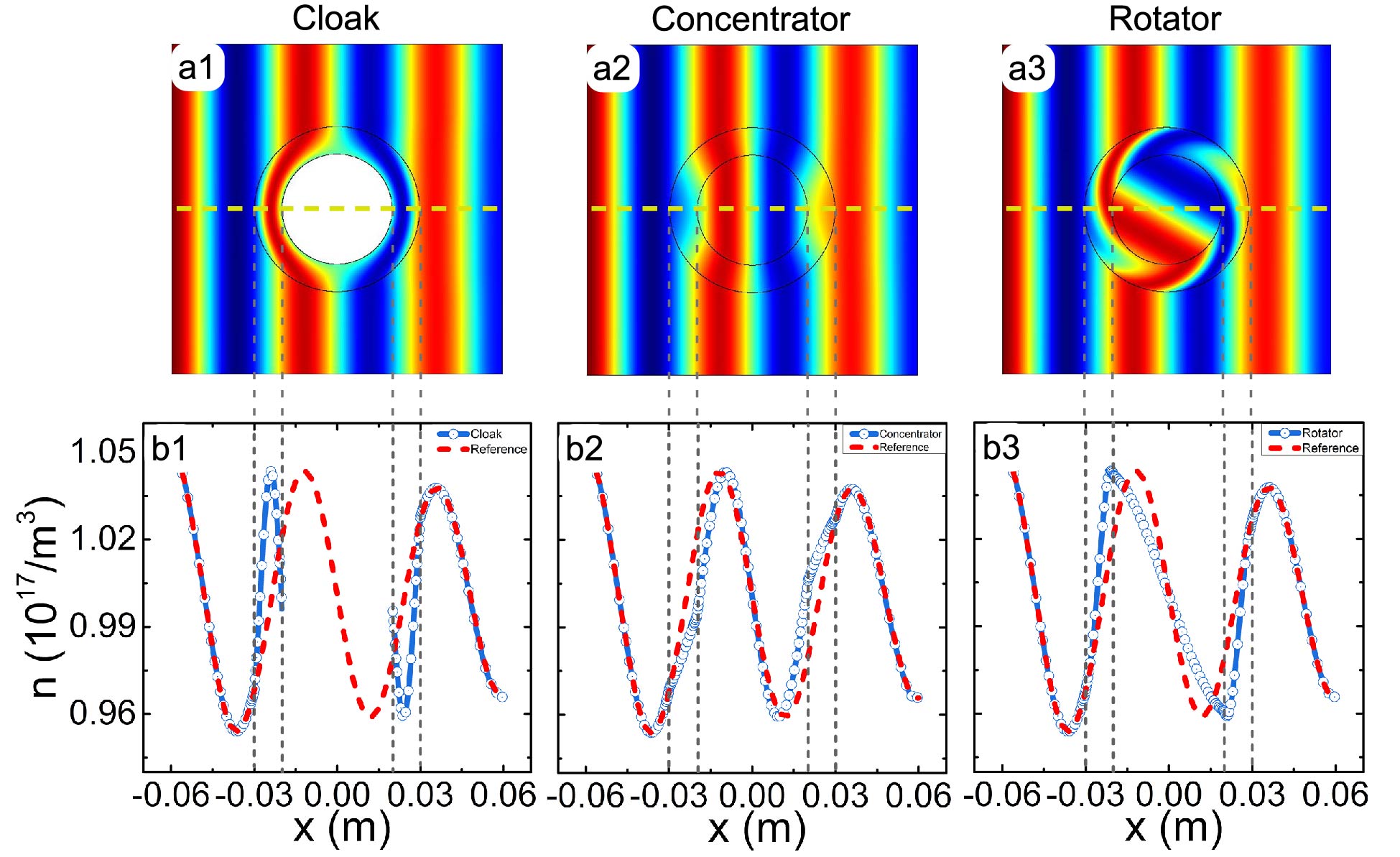}
	\caption{(a1)-(a3) Color mapping of density profiles at 40 s with a cloak, concentrator, and rotator, respectively. (b1)-(b3) Comparisons between density profiles in the pure background (reference) and those with a cloak, concentrator, and rotator, respectively. The grey dashed lines denote the position of the devices. The data are extracted along the yellow dashed line (y=0) in (a1)-(a3). Adapted from Ref.~\cite{Yfb-ZhangCPL22}.}
	\label{yfb-fig4}
\end{figure}

The models for the three devices have been presented in Fig.~\ref{c4fig2}. In order to visually depict the transient distribution of the plasma, a periodically fluctuating plasma source $n_b$ is chosen to impose on the left boundary of the square background. The specific expression for this is given by:
\begin{equation}
	n_b = n_1 \cos(\omega_0 t) + n_0,
\end{equation}
where $n_1 = 5.0 \times 10^{15}~\text{m}^{-3}$, $\omega_0 = \frac{2\pi}{10}~\text{s}^{-1}$, and $n_0 = 1.0 \times 10^{17}~\text{m}^{-3}$. The right side of the background is set as the outflow boundary, while the top and bottom boundaries are set as no flux. For the shield, its internal boundary should also be specifically set as no flux. In this context, the diffusion coefficient in the background is set as a constant $D_0$, and the electric field is set as a uniform electric field along the $x$-direction, represented by $\boldsymbol{E}_0$. Then all the parameters can be designed according to the above transformation rules, and the simulation results of cloaking, concentrating, and rotating are shown in Figs.~\ref{yfb-fig2}-\ref{yfb-fig4}, respectively.

Figure~\ref{yfb-fig2} illustrates the transient simulation of plasma transport under three conditions, namely, transporting in a pure background medium (set as the reference), in a background medium with a bare obstacle, and in a background medium with an obstacle covered by the cloak. The columns from left to right are screenshots of distributions of the plasma density at 10 s, 22 s, and 40 s, respectively. Due to the boundary condition of harmonically oscillating density, the plasma streams forward in a wave-like form. Moreover, the amplitude attenuation of the plasma flow reflected from the figures is caused by the diffusion, whose decay rate is codetermined by the oscillation frequency, diffusivity, and electric field. As a result, suitable values are carefully chosen to make the results more intuitive. The cloak designed with the transformation theory helps to cancel the scattering induced by the obstacle. Therefore, the density profiles of the background plasma keep nearly undisturbed, which shows the validity of the theory.

The transient simulation results for the concentrator and rotator are shown in Fig.~\ref{yfb-fig3}. The first row of snapshots shows the converging effect of the gradient of plasma density. In addition, as a determinant of the converging effect, a bigger ratio ($r_m/r_1$) would bring a higher converging effect. And the maximum ratio is $r_2/r_1$. For the rotator, the rotation of plasma flow appears in Figs.~\ref{yfb-fig2}(b1)-(b3). Linearly deflecting concentric circles in the virtual space can account for the gradual deflection of the density profiles. The target rotation angle in the core region is determined by $\theta_0$ in Eqs.~(\ref{c4ct3}). Particularly,  $\det\mathbf{J}=1$ for rotators helps to completely eliminate the disturbance to background plasma density.

To further explore the performance of the devices, the density values along a horizontal line (denoted by the yellow dashed lines in Fig.~\ref{yfb-fig4}) from the results at 40 s are extracted and the density distribution of functional devices is compared with that of reference. See Fig.~\ref{yfb-fig4}(b1)-(b3). Two regions should be remarked. One is the core region of the device, the other is the background. All the red dashed lines in Fig.~\ref{yfb-fig4}(b1)-(b3) denote the data of the reference, while the blue dotted lines represent the data of the cloak, concentrator, and rotator, respectively. In Fig.~\ref{yfb-fig4}(b1), it is clear that the data are well overlapped in the background, and the plasma is excluded well from the core region. Moreover, the relative difference in the plasma density in the background region was less than 0.15\%. In Fig.~\ref{yfb-fig4}(b2), the dotted line is indeed denser than the dashed line in the core region without being seriously dislocated in the background. And the relative difference was less than 0.13\%. In Fig.~\ref{yfb-fig4}(b3), the relative difference was less than 0.01\% which is far smaller than the value of the cloak or concentrator. As mentioned above, the accurate transformation form of Eq.~(\ref{c4mdt3}(a)) may account for this nearly zero difference. Overall, the simulation can confirm the feasibility and reliability of the theory.

\begin{figure}[!ht]
	\centering
	\includegraphics[width=.9\linewidth]{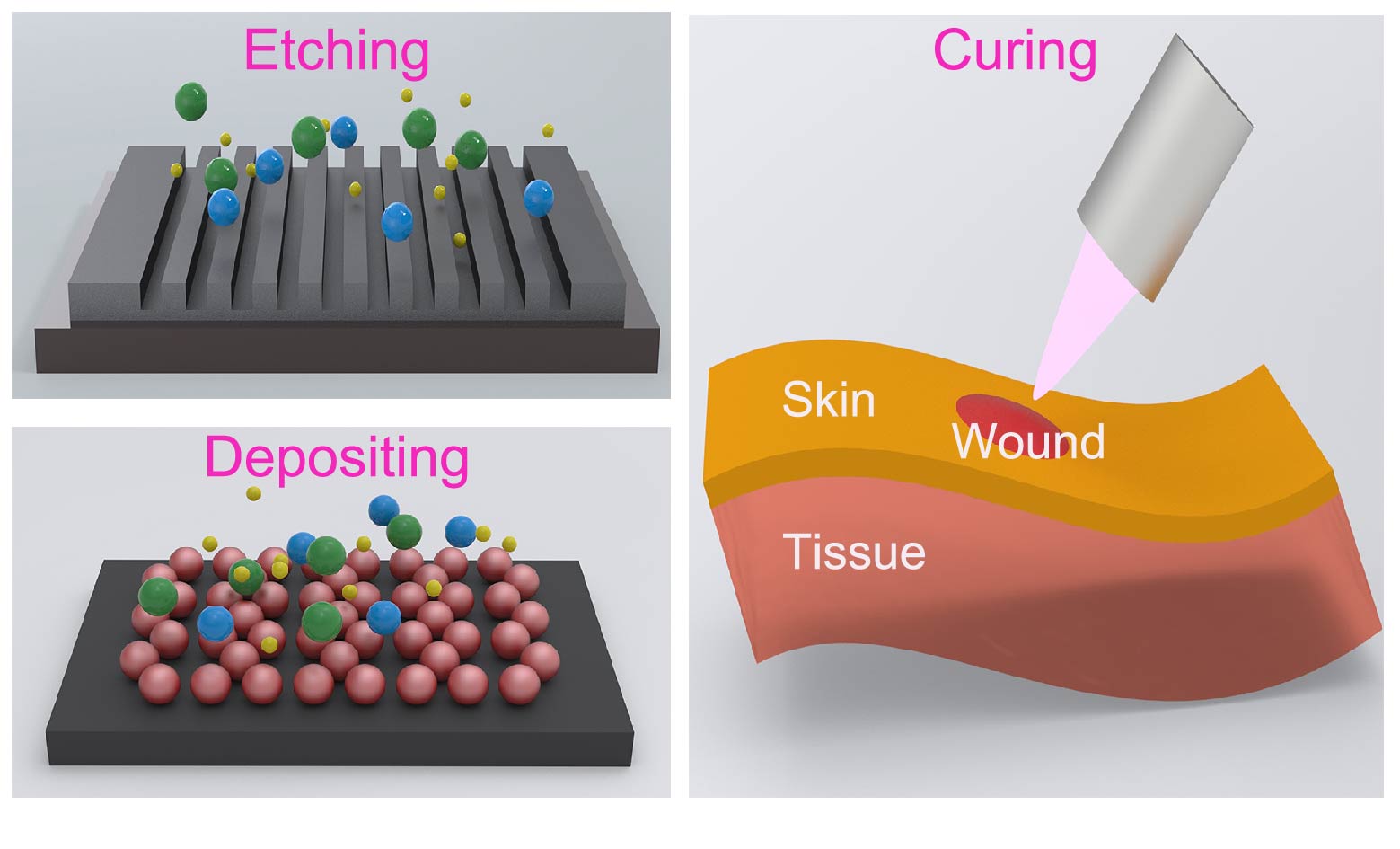}
	\caption{Applications of plasmas.}
	\label{yfb-fig5}
\end{figure}

The progression in plasma physics has paved the way for novel technologies and methods, finding cutting-edge applications in biomedicine, the crystal industry, and materials science~\cite{Yfb-LiangAEM18}. See Fig.~\ref{yfb-fig5}. We envisage several potential applications for devices crafted based on transformation theory. Consider the cloak, which has an isolated core region, as a prime candidate for safeguarding healthy tissue during plasma treatments of infected wounds. In catalyst development, a plasma flow convergence, characterized by a higher density of active particle clusters, augments the interaction between the plasma and the catalyst. This makes the concentrator an ideal tool to enhance catalytic performance. Furthermore, in aerospace, the concentrator might hold promise in elevating the efficiency of plasma-assisted engines. Beyond the uses already mentioned, the principles of coordinate transformation could be harnessed to achieve plasma separation or guidance, proving valuable for plasma etching or depositing. Additionally, transformation theory could contribute to the development of plasma metamaterials intended for electromagnetic wave manipulation~\cite{Yfb-RodPRAP21,Yfb-InaJAP21}.

Indeed, the proposed approach rooted in transformation theory holds merit. Even with the inherent challenges in manifesting the transformed diffusivities and electric fields, alternative techniques can be employed to achieve similar outcomes. Plenty of research has delved into customizing particle diffusivities. For instance, the scattering cancellation method allows for the creation of a bilayer diffusive cloak using two homogeneous materials~\cite{Yfb-ZhouCS21}. The complex diffusivity might be attainable through the effective medium theory~\cite{Yfb-ResSR16} or even machine learning techniques~\cite{Yfb-HuPRX20}. When it comes to electric field manipulation, insights from studies on electrostatic and magnetic cloaks could be illuminating~\cite{Yfb-NaraAM12,Yfb-LanSR15}.

While the future is ripe with possibilities, it also presents its fair share of challenges. Under broader circumstances, we must factor in the effects of magnetic fields and gas-phase reactions within the plasma. Truly exerting control over diffusivities and electric fields is a daunting task due to the intricate interplay between charged particles and electromagnetic fields. As such, it becomes paramount to incorporate alternative theoretical models or methodologies, such as the particle-in-cell/Monte Carlo collision model~\cite{Yfb-HuangIEEE15} or the nonequilibrium Green’s function approach~\cite{Yfb-YuFP15}. Additionally, plasma temperatures tend to fluctuate over time or space, especially in transient states, leading to a shift in transformation principles. In some scenarios, advection may play a role in plasma transport. Accounting for this term could diversify plasma regulation techniques. Furthermore, the emerging focus on spatiotemporal modulation in heat diffusion could provide new insights into plasma physics~\cite{Yfb-XingCPL21}. In conclusion, refining the transformation theory for plasmas requires heightened focus, research, and dedication.

\section{Potential Impacts for Novel Physics}
The realm of plasma transport transformation, being in its nascent stage, harbors a vast expanse of uncharted physics awaiting exploration. Among the intriguing avenues is the challenge of nonreciprocal plasma transport. Given the consistent display of space-inversion symmetry in diffusion equations, pinpointing nonreciprocal mechanisms within plasma transport proves daunting. An effective approach might involve the application of spatiotemporal modulations or nonlinear parameters, as found in the dynamics of heat and particle diffusion~\cite{Yfb-Torrent2018,Yfb-Zang2019,Yfb-Guo2019,Yfb-Camacho2020a,Yfb-Li2022}. Another captivating area is topological plasma transport. With lessons drawn from well-established mechanisms in topological particle and heat diffusion~\cite{Yfb-GongPRX18,Yfb-KawabataPRX19,Yfb-KlitzingPRL80,Yfb-Klitzing05,Yfb-HasanRMP10,Yfb-QiRMP11,Yfb-LuNP14,Yfb-OzawaRMP19,Yfb-YangPRL15,Yfb-XueNRM22,Yfb-HuberNP16}, researchers could embark on a variety of experimental strategies. For instance, formulating a plasma analog of the Su-Schrieffer-Heeger model might shed light on the intricate interplay between bulk-boundary and the emergence of edge states~\cite{Yfb-ParkerJPP21}. Additionally, the exploration into the presence of exceptional points and geometric phases promises rich insights. In a linear context, the use of Chern numbers can act as descriptors for the unconventional topologies found in plasma. Furthermore, the introduction of elements such as flow shear or a density gradient could lead to non-Hermiticity in systems, thereby highlighting plasma as a key medium for delving into the complexities of non-Hermitian physics. Given these vast possibilities, we anticipate a slew of groundbreaking discoveries in this domain in the near future.

\section{Conclusion}

In this review, a toy model, specifically the diffusion-migration model, has been utilized to elucidate plasma transport. This investigation has highlighted the practicality of the transformation theory. It was revealed that the transformed diffusion-migration equation maintains form-invariance at steady states but deviates at transient states. However, it was also illustrated that by setting minimal diffusivities, the transformed transient equation can attain an approximate form-invariance. Based on this, three conceptual model devices were conceptualized, serving as plasma cloaks, concentrators, or rotators for transient plasma transport. Such findings can potentially expand methods for manipulating plasma flow and offer applications across multiple sectors, including medicine and aerospace. In conclusion, controlling plasma may serve as a nexus between diffusion and wave metamaterials. Although there are evident differences between diffusion and wave metamaterials, simultaneous control over diffusion and wave propagation signifies a notable progression. Plasma transport predominantly follows a diffusion process, but plasmas are frequently harnessed to regulate electromagnetic waves due to their distinctive permittivity. As such, the convergence of diffusion and wave propagation within a singular system can be a precursor to the emergence of new physics.

\section*{Acknowledgments}
We acknowledge the financial support provided by the National Natural Science Foundation of China under Grants No. 12035004 and No. 12320101004, the Science and Technology Commission of Shanghai Municipality under Grant No. 20JC1414700, and the Innovation Program of Shanghai Municipal Education Commission under Grant No. 2023ZKZD06.


\begin{thebibliography}{99}
\bibliographystyle{apsrev4-1}

\bibitem{Yfb-Ve} Veselago, V.G.: The electrodynamics of substances with simultaneously negative
values of $\rm{\epsilon}$ and $\rm{\mu}$. Sov. Phys. Usp. {\bf 10}, 509-514 (1968)

\bibitem{Yfb-PendryPRL96} Pendry, J.B., Holden, A.J., Stewart, W.J., Youngs, I.: Extremely low frequency plasmons in metallic mesostructures. Phys. Rev. Lett. {\bf 76}, 4773-4776 (1996)

\bibitem{Yfb-PendryPRL99} Pendry, J.B., Holden, A.J., Robbins, D.J., Stewart, W.J.: Magnetism from conductors
and enhanced nonlinear phenomena. IEEE Trans. Microw. Theory Tech. {\bf 47}, 2075-2084 (1999)

\bibitem{Yfb-LeoSci06} Leonhardt, U.: Optical conformal mapping. Science {\bf 312}, 1777-1780 (2006)

\bibitem{Yfb-PenSci06} Pendry, J.B., Schurig, D., Smith, D.R.: Controlling electromagnetic fields. Science {\bf 312}, 1780-1782 (2006)

\bibitem{Yfb-ChenNM10} Chen, H., Chan, C.T., Sheng, P.: Transformation optics and metamaterials. Nat. Mater. {\bf 9}, 387-396 (2010)

\bibitem{Yfb-ZheNM12} Zheludev, N.I., Kivshar, Y.S.: From metamaterials to metadevices. Nat. Mater. {\bf 11}, 917-924 (2012)

\bibitem{Yfb-KadicRPP13} Kadic, M., B\"{u}ckmann, T., Schittny, R., Wegener, M.: Metamaterials beyond electromagnetism. Rep. Prog. Phys. {\bf 76}, 126501 (2013)

\bibitem{Yfb-FuPQE19} Fu, X., Cui, T.J.: Recent progress on metamaterials: From effective medium model
to real-time information processing system. Prog. Quantum Electron. {\bf 67}, 100223 (2019)


\bibitem{Yfb-FanAPL2008} Fan, C.Z., Gao, Y., Huang J.P.: Shaped graded materials with an apparent negative thermal conductivity. Appl. Phys. Lett. {\bf 92}, 251907 (2008)

\bibitem{Yfb-ZhangNRP23} Zhang, Z.R., Xu, L.J., Qu, T., Lei, M., Lin, Z.K., Ouyang, X.P., Jiang, J.-H., Huang, J.P.: Diffusion metamaterials. Nat. Rev. Phys. {\bf 5}, 218-235 (2023)
\bibitem{Yfb-rmp} Yang, F.B., Zhang, Z.R., Xu, L.J., Liu, Z.F., Jin, P., Zhuang, P.F., Lei, M., Liu, J.R., Jiang, J.-H., Ouyang, X.P., Marchesoni, F., Huang, J.P.: Controlling mass and energy diffusion with metamaterials. Rev. Mod. Phys. accepted (2023)
\bibitem{Yfb-LiuEPJAP2009} Liu, B., Huang, J.P.: Acoustically conceal an object with hearing. Eur. Phys. J. Appl. Phys. {\bf 48}, 20501 (2009)

\bibitem{Yfb-LiuCTP10} Liu, B., Huang, J.P.: Noise shielding using acoustic metamaterials. Commun. Theor. Phys. {\bf 53}, 560-564 (2010)


\bibitem{Yfb-QiuAIP Adv.15}Zhu, N.Q., Shen, X.Y., Huang, J.P.: Converting the patterns of local heat flux via thermal illusion device. AIP Adv. {\bf 5}, 053401 (2015)



\bibitem{Yfb-Tan15} Shen, X.Y., Chen, Y.X., Huang, J.P.: Thermal magnifier and minifier. Commun. Theor. Phys. {\bf 65}, 375-380 (2016)


\bibitem{Yfb-ShenPRL16} Shen, X.Y., Li, Y., Jiang C.R., Huang, J.P.: Temperature trapping: Energy-free maintenance of constant temperatures as ambient temperature gradients change. Phys. Rev. Lett. {\bf 117}, 055501 (2016)

\bibitem{Yfb-HuangPB17} Huang, J.Y., Shen, X.Y., Jiang, C.R., Wu, Z.H., Huang, J.P.: Thermal expander. Physica B {\bf 518}, 56-60 (2017)



\bibitem{Yfb-JiIJMPB18} Ji, Q., Shen, X.Y., Huang, J.P.: Transformation thermotics: Thermal metamaterials and their applications. Int. J. Mod. Phys. B {\bf 32}, 1840004 (2018)

\bibitem{Yfb-XuEPJB18} Xu, L.J., Jiang, C.R., Huang, J.P.: Heat-source transformation thermotics: From boundary-independent conduction to all-directional replication. Eur. Phys. J. B {\bf 91}, 166 (2018)

\bibitem{Yfb-XuJAP18} Xu, L.J., Wang, R.Z., Huang, J.P.: Camouflage thermotics: A cavity without disturbing heat signatures outside. J. Appl. Phys. {\bf 123}, 245111 (2018)

\bibitem{Yfb-XuPLA18} Xu, L.J., Huang, J.P.: A transformation theory for camouflaging arbitrary heat sources. Phys. Lett. A {\bf 382}, 3313-3316 (2018)



\bibitem{Yfb-HuangPP18} Huang, J.P.: Thermal metamaterial: Geometric structure, working mechanism, and novel function. Prog. Phys. {\bf 38}, 219-238 (2018)



\bibitem{Yfb-HuangESEE19} Huang, J.P.: Thermal metamaterials make it possible to control the flow of heat at will. ES Energy Environ. {\bf 6}, 1-3 (2019)

\bibitem{Yfb-Huang20} Huang, J.P.: Theoretical Thermotics: Transformation Thermotics and Extended Theories for Thermal Metamaterials. Springer, Singapore (2020)

\bibitem{Yfb-XuPRAP20} Xu, L.J., Dai, G.L., Huang, J.P.: Transformation multithermotics: Controlling radiation and conduction simultaneously. Phys. Rev. Appl. {\bf 13}, 024063 (2020)

\bibitem{Yfb-XuESEE20} Xu, L.J., Yang, S., Dai, G.L., Huang, J.P.: Transformation omnithermotics: Simultaneous manipulation of three basic modes of heat transfer. ES Energy Environ. {\bf 7}, 65-70 (2020)

\bibitem{Yfb-HuangESEE20} Huang, J.P.: Thermal metamaterials make it possible to control the flow of heat at will. ES Energy Environ. {\bf 7}, 1-3 (2020)

\bibitem{Yfb-XuIJHMT20} Xu, L.J., Huang, J.P.: Controlling thermal waves with transformation complex thermotics. Int. J. Heat Mass Transf. {\bf 159}, 120133 (2020)

\bibitem{Yfb-HuangPhysics20} Huang, J.P.: On theoretical thermotics (Invited review in Chinese). Physics {\bf 49}, 493-496 (2020)

\bibitem{Yfb-WangiScience20} Wang, J., Dai, G.L., Huang, J.P.: Thermal metamaterial: Fundamental, application, and outlook. iScience {\bf 23}, 101637 (2020)





\bibitem{Yfb-YangPR21} Yang, S., Wang, J., Dai, G.L., Yang, F.B., Huang, J.P.: Controlling macroscopic heat transfer with thermal metamaterials: Theory, experiment and application. Phys. Rep. {\bf 908}, 1-65 (2021)

\bibitem{Yfb-LeiEPL21} Lei, M., Wang, J., Dai, G.L., Tan, P., Huang, J.P.: Temperature-dependent transformation multiphysics and ambient-adaptive multiphysical metamaterials. EPL {\bf 135}, 54003 (2021)

\bibitem{Yfb-ZhangATS22} Zhang, Z.R., Xu, L.J., Huang, J.P.: Controlling Chemical Waves by Transforming Transient Mass Transfer. Adv. Theory Simul. {\bf 5}, 2100375 (2022)

\bibitem{Yfb-WangCPB22} Wang, B., Huang, J.P.: Hydrodynamic metamaterials for flow manipulation: Functions and prospects. Chin. Phys. B {\bf 31}, 098101 (2022)

\bibitem{Yfb-YangPRAP22} Yang, F.B., Xu, L.J., Wang, J., Huang, J.P.: Transformation Theory for Spatiotemporal Metamaterials. Phys. Rev. Appl. {\bf 18}, 034080 (2022)

\bibitem{Yfb-XuBook23} Xu, L.J., Huang, J.P.: Transformation Thermotics and Extended Theories: Inside and Outside Metamaterials. Springer, 2023

\bibitem{Yfb-YangPRA23} Yang, F.B., Jin, P., Lei, M., Dai, G.L., Wang, J., Huang, J.P.: Space-time thermal binary coding by spatiotemporally modulated metashell. Phys. Rev. Appl. {\bf 19}, 054096 (2023)

\bibitem{Yfb-DaiPR23} Dai, G.L., Yang, F.B., Xu, L.J., Huang, J.P.: Diffusive pseudo-conformal mapping: Anisotropy-free transformation thermal media with perfect interface matching. Chaos, Solitons \& Fractals. {\bf 174}, 113849 (2023)

\bibitem{Yfb-Tan20} Shen, X.Y., Huang, J.P.: Transformation thermotics: Thermal metamaterials and their applications. Acta Physica Sinica. {\bf 65}, 178103 (2016)


\bibitem{Yfb-FanCTP10} Fan, C.Z., Gao, Y.H., Gao, Y., Huang, J.P.: Apparently negative electric polarization in shaped graded dielectric metamaterials. Commun. Theor. Phys. {\bf 53}, 913-919 (2010)

\bibitem{Yfb-ZhaoFOP12} Zhao, L., Liu, B., Gao, Y.H., Zhao, Y.J., Huang, J.P.: Enhanced scattering of acoustic waves at interfaces. Front. Phys. {\bf 7}, 319-323 (2012)

\bibitem{Yfb-ChenEPJAP15} Chen, Y.X., Shen, X.Y., Huang, J.P.: Engineering the accurate distortion of an object's temperature-distribution signature. Eur. Phys. J. Appl. Phys. {\bf 70}, 20901 (2015)

\bibitem{Yfb-ShenAPL16} Shen, X.Y., Jiang C.R., Li, Y., Huang, J.P.: Thermal metamaterial for convergent transfer of conductive heat with high efficiency. Appl. Phys. Lett. {\bf 109}, 201906 (2016)

\bibitem{Yfb-YangAPL17} Yang, S., Xu, L.J., Wang, R.Z., Huang, J.P.: Full control of heat transfer in single-particle structural materials. Appl. Phys. Lett. {\bf 111}, 121908 (2017)

\bibitem{Yfb-XuEPJB17} Xu, L.J., Jiang, C.R., Shang, J., Wang R.Z., Huang, J.P: Periodic composites: Quasi-uniform heat conduction, Janus thermal illusion, and illusion thermal diodes. Eur. Phys. J. B {\bf 90}, 221 (2017)

\bibitem{Yfb-WangJAP17} Wang, R.Z., Xu, L.J., Huang, J.P.: Thermal imitators with single directional invisibility. J. Appl. Phys. {\bf 122}, 215107 (2017)

\bibitem{Yfb-ShangIJHMT18} Shang, J., Wang, R.Z., Xin, C., Dai, G.L., Huang, J.P.: Macroscopic networks of thermal conduction: Failure tolerance and switching processes. Int. J. Heat Mass Transf. {\bf 121}, 321-328 (2018)

\bibitem{Yfb-JiCTP18} Ji, Q., Huang, J.P.: Controlling thermal conduction by graded materials. Commun. Theor. Phys. {\bf 69}, 434-440 (2018)

\bibitem{Yfb-DaiEPJB18} Dai, G.L., Shang, J., Wang, R.Z., Huang, J.P.: Nonlinear thermotics: Nonlinearity enhancement and harmonic generation in thermal metasurfaces. Eur. Phys. J. B {\bf 91}, 59 (2018)



\bibitem{Yfb-ShangJHT18} Shang, J., Jiang, C.R., Xu, L.J., Huang, J.P.: Many-particle thermal invisibility and diode from effective media. J. Heat Transf.-Trans. ASME {\bf 140}, 092004 (2018)

\bibitem{Yfb-WangIJTS18} Wang, R.Z., Shang, J., Huang, J.P.: Design and realization of thermal camouflage with many-particle systems. Int. J. Therm. Sci. {\bf 131}, 14-19 (2018)

\bibitem{Yfb-XuPRE18} Xu, L.J., Yang, S., Huang, J.P.: Designing the effective thermal conductivity of materials of core-shell structure: Theory and simulation. Phys. Rev. E {\bf 98}, 052128 (2018)

\bibitem{Yfb-WangPRA19} Wang, J., Shang, J., Huang, J.P.: Negative energy consumption of thermostats at ambient temperature: Electricity generation with zero energy maintenance. Phys. Rev. Appl. {\bf 11}, 024053 (2019)

\bibitem{Yfb-XuPRA19a} Xu, L.J., Yang, S., Huang, J.P.: Thermal transparency induced by periodic interparticle interaction. Phys. Rev. Appl. {\bf 11}, 034056 (2019)

\bibitem{Yfb-YangPRE19} Yang, S., Xu, L.J., Huang, J.P.: Metathermotics: Nonlinear thermal responses of core-shell metamaterials. Phys. Rev. E {\bf 99}, 042144 (2019)

\bibitem{Yfb-ZhouESEE19} Zhou, Z.Y., Shen, X.Y., Fang, C.C., Huang, J.P.: Programmable thermal metamaterials based on optomechanical systems. ES Energy Environ. {\bf 6}, 85-91 (2019)

\bibitem{Yfb-DaiIJHMT20} Dai, G.L., Huang, J.P.: Nonlinear thermal conductivity of periodic composites. Int. J. Heat Mass Transf. {\bf 147}, 118917 (2020)

\bibitem{Yfb-WangPRE20} Wang, J., Dai, G.L., Yang, F.B., Huang, J.P.: Designing bistability or multistability in macroscopic diffusive systems. Phys. Rev. E {\bf 101}, 022119 (2020)

\bibitem{Yfb-XuEPJB20} Xu, L.J., Zhao, X.T., Zhang, Y.P., Huang, J.P.: Tailoring dipole effects for achieving thermal and electrical invisibility simultaneously. Eur. Phys. J. B {\bf 93}, 101 (2020)

\bibitem{Yfb-WangPRAP20} Wang, J., Yang, F.B., Xu, L.J., Huang, J.P.: Omnithermal restructurable metasurfaces for both infrared-light illusion and visible-light similarity. Phys. Rev. Appl. {\bf 14}, 014008 (2020)

\bibitem{Yfb-XuAPL20} Xu, L.J., Huang, J.P.: Thermal convection-diffusion crystal for prohibition and modulation of wave-like temperature profiles. Appl. Phys. Lett. {\bf 117}, 011905 (2020)

\bibitem{Yfb-XuCPLEL20} Xu, L.J., Huang, J.P.: Negative thermal transport in conduction and advection. Chin. Phys. Lett. (Express Letter) {\bf 37}, 080502 (2020)

\bibitem{Yfb-WangICHMT20} Wang, B., Shih, T.M., Huang, J.P.: Enhancing and attenuating heat transfer characteristics for circulating flows of nanofluids within rectangular enclosures. Int. Comm. Heat Mass Transf. {\bf 117}, 104800 (2020)



\bibitem{Yfb-DaiJNU21} Dai, G.L., Huang, J.P.: Nonlinear thermotics: Designing thermal metamaterials with temperature response (Invited review in Chinese). Journal of Nantong University (Natural Science Edition) {\bf 20}, 1-18 (2021)

\bibitem{Yfb-XuEPL21} Xu, L.J., Yang, S., Huang, J.P.: Controlling thermal waves of conduction and convection. Europhys. Lett. {\bf 133}, 20006 (2021)

\bibitem{Yfb-XuPRE21} Xu, L.J., Huang, J.P., Ouyang, X.P.: Tunable thermal wave nonreciprocity by spatiotemporal modulation. Phys. Rev. E {\bf 103}, 032128 (2021)

\bibitem{Yfb-ZhangTSEP21} Zhang, Z.R., Xu, L.J., Ouyang, X.P., Huang, J.P.: Guiding temperature waves with graded metamaterials. Therm. Sci. Eng. Prog. {\bf 23}, 100926 (2021)

\bibitem{Yfb-TianIJHMT21} Tian, B.Y., Wang, J., Dai, G.L., Ouyang, X.P., Huang, J.P.: Thermal metadevices with geometrically anisotropic heterogeneous composites. Int. J. Heat Mass Transfer {\bf 174}, 121312 (2021)

\bibitem{Yfb-XuAPL21} Xu, L.J., Huang, J.P., Ouyang, X.P.: Nonreciprocity and isolation induced by an angular momentum bias in convection-diffusion systems. Appl. Phys. Lett. {\bf 118}, 22190 (2021)

\bibitem{Yfb-XuEPL21-2} Xu, L.J., Huang, J.P.: Robust one-way edge state in convection-diffusion systems. EPL {\bf 134}, 60001 (2021)

\bibitem{Yfb-HuangAMT22} Huang, T.Q., Yang, F.B., Wang, T., Wang, J., Li, Y.W., Huang, J.P., Chen, M., Wu, L.M.: Ladder-structured boron nitride nanosheet skeleton in flexible polymer films for superior thermal conductivity. Appl. Mater. Today {\bf 26}, 101299 (2022)


\bibitem{Yfb-XuPRL22} Xu, L.J., Xu, G.Q., Huang, J.P., Qiu, C-W.: Diffusive Fizeau Drag in Spatiotemporal Thermal Metamaterials. Phys. Rev. Lett. {\bf 128}, 145901 (2022)

\bibitem{Yfb-LinSCPMA22} Lin, W.Y., Zhang, H.Y., Kalcheim, Y., Zhou, X.C., Yang, F.B., Shi, Y., Feng, Y., Wang, Y.H., Huang, J.P., Schuller, I.K., Zhou, X.D., Shen, J.: Direct visualization of percolating metal-insulator transition in V2O3 using scanning microwave impedance microscopy. Sci. China Phys. Mech. Astron. {\bf 65}, 297411 (2022)

\bibitem{Yfb-ZhuangPRE22} Zhuang, P.F., Wang, J., Yang, S., Huang, J.P.: Nonlinear thermal responses in geometrically anisotropic metamaterials. Phys. Rev. E. {\bf 106}, 044203 (2022)

\bibitem{Yfb-XuPRL22-2} Xu, L.J., Xu, G.Q., Li, J.X., Li, Y., Huang, J.P., Qiu, C.-W.: Thermal Willis coupling in spatiotemporal diffusive metamaterials. Phys. Rev. Lett. {\bf 129}, 155901 (2022)

\bibitem{Yfb-ZhouEPL23} Zhou, X.C., Lin, W.Y., Yang, F.B., Zhou, X.D., Shen, J., Huang, J.P.: Effective medium theory with hybrid impacts of phase symmetry and asymmetry for analyzing phase transition behavior. Eur. Phys. Lett. {\bf 141}, 16001 (2023)

\bibitem{Yfb-LeiIJHMT23} Lei, M., Jiang, C.R., Yang, F.B., Wang, J., Huang, J.P.: Programmable all-thermal encoding with metamaterials. Int. J. Heat Mass Transfer. {\bf 207}, 124033 (2023)

\bibitem{Yfb-ZhuangIJMSD23} Zhuang, P.F., Huang, J.P.: Multiple control of thermoelectric dual-function metamaterials. Int. J. Mech. Sys. Dyna {\bf 3}, 127-135 (2023)

\bibitem{Yfb-XuNSR23} Xu, L.J, Liu, J.R., Jin, P., Xu, G.Q., Li, J.X., Ouyang, X.P., Li, Y., Qiu, C.-W., Huang, J.P.: Black-hole-inspired thermal trapping with graded heat-conduction metadevices. Nat. Sci. Rev. {\bf 10}, nwac159 (2023)

\bibitem{Yfb-LeiMTP23} Lei, M., Xu, L.J., Huang, J.P.: Spatiotemporal multiphysics metamaterials with continuously adjustable functions. Mat. Today. Phys. {\bf 34}, 101057 (2023)

\bibitem{Yfb-XuPANS23} Xu, L.J., Liu, J,R., Xu, G.Q., Huang, J.P., Qiu, C.-W.: Giant, magnet-free, and room-temperature Hall-like heat transfer. Proc. Natl. Acad. Sci. U.S.A. {\bf 120}, e2305755120 (2023)

\bibitem{Yfb-ZhangCPL23} Zhang, C.X, Li, T.J., Xu, L.J., Huang, J.P.: Dust-induced regulation of thermal radiation in water droplets. Chin. Phys. Lett. {\bf 40}, 054401 (2023) 

\bibitem{Yfb-click} Wang, C.M., Jin, P., Yang, F.B., Xu, L.J., Huang, J.P.: Click metamaterials: Fast acquisition of thermal conductivity and functionality diversities. Preprint at https://doi.org/10.48550/arXiv.2308.16057 (2023)

\bibitem{Yfb-research} Jin, P., Liu, J.R., Yang, F.B., Marchesoni, F., Jiang, J.-H., Huang, J.P.: In-situ simulation of thermal reality. Research {\bf 6}, 0222 (2023)

\bibitem{Yfb-nc} Zhou, X.C., Xu, X., Huang, J.P.: Adaptive multi-temperature control for transport and storage containers enabled by phase change materials. Nat. Commun. {\bf 14}, 5449 (2023)


\bibitem{Yfb-XuEPJB19} Xu, L.J., Huang, J.P.: Electrostatic chameleons: Theory of intelligent metashells with adaptive response to inside objects. Eur. Phys. J. B {\bf 92}, 53 (2019)

\bibitem{Yfb-XuEPL19} Xu, L.J., Huang, J.P.: Magnetostatic chameleonlike metashells with negative permeabilities. Europhys. Lett. {\bf 125}, 64001 (2019)

\bibitem{Yfb-XuPRA19} Xu, L.J., Yang, S., Huang, J.P.: Passive metashells with adaptive thermal conductivities: Chameleonlike behavior and its origin. Phys. Rev. Appl. {\bf 11}, 054071 (2019)

\bibitem{Yfb-YangEPL19} Yang, S., Xu, L.J., Huang, J.P.: Intelligence thermotics: Correlated self-fixing behavior of thermal metamaterials. Europhys. Lett. {\bf 126}, 54001 (2019)



\bibitem{Yfb-XuEPJB19a} Xu, L.J., Yang, S., Huang, J.P.: Thermal illusion with the concept of equivalent thermal dipole. Eur. Phys. J. B {\bf 92}, 264 (2019)   

\bibitem{Yfb-YangEPL19a} Yang, S., Xu, L.J., Huang, J.P.: Two exact schemes to realize thermal chameleonlike metashells. EPL {\bf 128}, 34002 (2019) 

\bibitem{Yfb-XuPRE19} Xu, L.J., Yang, S., Huang, J.P.: Dipole-assisted thermotics: Experimental demonstration of dipole-driven thermal invisibility. Phys. Rev. E {\bf 100}, 062108 (2019)  

\bibitem{Yfb-XuSCPMA20} Xu, L.J., Huang, J.P.: Chameleonlike metashells in microfluidics: A passive approach to adaptive responses. Sci. China-Phys. Mech. Astron. {\bf 63}, 228711 (2020)  

\bibitem{Yfb-SuEPL20} Su, C., Xu, L.J., Huang, J.P.: Nonlinear thermal conductivities of core-shell metamaterials: Rigorous theory and intelligent application. EPL {\bf 130}, 34001 (2020)  











\bibitem{Yfb-QuEPL21} Qu, T., Wang, J., Huang, J.P.: Manipulating thermoelectric fields with bilayer schemes beyond Laplacian metamaterials. EPL {\bf 135}, 54004 (2021)




\bibitem{Yfb-JinPNAS23} Jin, P., Liu, J.R., Xu, L.J., Wang, J., Ouyang, X.P., Jiang, J.-H., Huang, J.P.: Tunable liquid-solid hybrid thermal metamaterials with a topology transition. Proc. Natl. Acad. Sci. U.S.A. {\bf 120}, e2217068120 (2023)

\bibitem{Yfb-ZhangPRA23} Zhang, Z.R., Yang, F.B., Huang, J.P.: Intelligent chameleonlike metashells for mass diffusion. Phys. Rev. Appl. {\bf 19}, 024009 (2023)


\bibitem{Yfb-LiuJAP21} Liu, B., Xu, L.J., Huang, J.P.: Thermal transparency with periodic particle distribution: A machine learning approach. J. Appl. Phys. {\bf 129}, 065101 (2021)



\bibitem{Yfb-LiuJAP21-2} Liu, B., Xu, L.J., Huang, J.P.: Reinforcement learning approach to thermal transparency with particles in periodic lattices. J. Appl. Phys. {\bf 130}, 045103 (2021)

\bibitem{Yfb-ZhangPRD22} Zhang, C.X., Li, T.J., Jin, P., Yuan, Y., Ouyang, X.P., Marchesoni, F., Huang, J.P.: Extracting stellar emissivity via a machine learning analysis of MSX and LAMOST catalog data. Phys. Rev. D. {\bf 106}, 123035 (2022)
\bibitem{Yfb-am} Jin, P., Xu, L., Xu, G., Li, J., Qiu, C.-W., Huang, J.P.: Machine-learning-assisted environment-adaptive thermal metamaterials. Preprint at https://arxiv.org/abs/2301.04523 (2023)


\bibitem{Yfb-ChenAPL2008} Chen, T.Y., Weng, C.N., Chen, J.S.: Cloak for curvilinearly anisotropic media in
conduction. Appl. Phys. Lett. {\bf 93}, 114103 (2008)

\bibitem{Yfb-GaoJAP2009} Gao, Y., Huang, J.P., Yu, K.W.: Multifrequency cloak with multishell by using transformation medium. J. Appl. Phys. {\bf 105}, 124505 (2009)

\bibitem{Yfb-LiJAP10} Li, J.Y., Gao, Y., Huang, J.P.: A bifunctional cloak using transformation media. J. Appl. Phys. {\bf 108}, 074504 (2010)

\bibitem{Yfb-SuFOP11} Su, Q., Liu, B., Huang, J.P.: Remote acoustic cloaks. Front. Phys. {\bf 6}, 65 (2011)

\bibitem{Yfb-QiuEPL13} Gao, Y., Huang, J.P.: Unconventional thermal cloak hiding an object outside the cloak. EPL {\bf 104}, 44001 (2013)



\bibitem{Yfb-QiuIJHT14}Shen, X.Y., Huang, J.P.: Thermally hiding an object inside a cloak with feeling. Int. J. Heat Mass Transfer {\bf 78}, 1-6 (2014)


\bibitem{Yfb-Tan16} Li, Y., Shen, X.Y., Wu, Z.H., Huang, J.Y., Chen, Y.X., Ni, Y.S., Huang, J.P.: Temperature-dependent transformation thermotics: From switchable thermal cloaks to macroscopic thermal diodes. Phys. Rev. Lett. {\bf 115}, 195503 (2015)



\bibitem{Yfb-YangJAP19} Yang, S., Xu, L.J., Huang, J.P.: Thermal magnifier and external cloak in ternary component structure. J. Appl. Phys. {\bf 125}, 055103 (2019)

\bibitem{Yfb-XuCPL20} Xu, L.J., Huang, J.P.: Active thermal wave cloak. Chin. Phys. Lett. {\bf 37}, 120501 (2020)


\bibitem{Yfb-WangATE21} Wang, B., Shih, T.M., Huang, J.P.: Transformation heat transfer and thermo-hydrodynamic cloaks for creeping flows: Manipulating heat fluxes and fluid flows simultaneously. Appl. Therm. Eng. {\bf 190}, 116726 (2021)


\bibitem{Yfb-WangPRAP21} Wang, B., Shih, T.M., Xu, L.J., Dai, G.L., Huang, J.P.: Intangible hydrodynamic cloaks for convective flows. Phys. Rev. Appl. {\bf 15}, 034014 (2021)


\bibitem{Yfb-XuIJHMT21} Xu, L.J., Wang, J., Dai, G.L., Yang, S., Yang, F.B., Wang, G., Huang, J.P.: Geometric phase, effective conductivity enhancement, and invisibility cloak in thermal convection-conduction. Int. J. Heat Mass Transf. {\bf 165}, 120659 (2021)



\bibitem{Yfb-XuPRAP19} Xu, L.J., Huang, J.P.: Metamaterials for manipulating thermal radiation: Transparency, cloak, and expander. Phys. Rev. Appl. {\bf 12}, 044048 (2019)  

\bibitem{Yfb-XuEPL20a} Xu, L.J., Yang, S., Huang, J.P.: Effectively infinite thermal conductivity and zero-index thermal cloak. EPL {\bf 131}, 24002 (2020)   

\bibitem{Yfb-YangJAP20} Yang, S., Xu, L.J., Dai, G.L., Huang, J.P.: Omnithermal metamaterials switchable between transparency and cloaking. J. Appl. Phys. {\bf 128}, 095102 (2020)

\bibitem{Yfb-XuPRE20} Xu, L.J., Dai, G.L., Wang, G., Huang, J.P.: Geometric phase and bilayer cloak in macroscopic particle-diffusion systems. Phys. Rev. E {\bf 102}, 032140 (2020)



\bibitem{Yfb-DaiPRAP22} Dai, G.L., Zhou, Y.H., Wang, J., Yang, F.B., Qu, T., Huang, J.P.: Convective Cloak in Hele-Shaw Cells with Bilayer Structures: Hiding Objects from Heat and Fluid Motion Simultaneously. Phys. Rev. Appl. {\bf 17}, 044006 (2022)

\bibitem{Yfb-YaoISci22} Yao, N.Z., Wang, H., Wang, B., Wang, X.S., Huang, J.P.: Convective thermal cloaks with homogeneous and isotropic parameters and drag-free characteristics for viscous potential flows. iScience. {\bf 25}, 105461 (2022)



\bibitem{Yfb-Tan4} Shen, X.Y., Li, Y., Jiang, C.R., Ni, Y.S. Huang, J.P.: Thermal cloak-concentrator. Appl. Phys. Lett. {\bf 109}, 031907 (2016)

\bibitem{Yfb-Tan11} Li, Y., Shen, X.Y., Huang, J.P., Ni, Y.S.: Temperature-dependent transformation thermotics for unsteady states: Switchable concentrator for transient heat flow. Phys. Lett. A {\bf 380}, 1641-1647 (2016)

\bibitem{Yfb-DaiPRE18} Dai, G.L., Shang, J., Huang, J.P.: Theory of transformation thermal convection for creeping flow in porous media: Cloaking, concentrating, and camouflage. Phys. Rev. E {\bf 97}, 022129 (2018)



\bibitem{Yfb-WangJAP18} Wang, R.Z., Xu, L.J., Ji, Q., Huang, J.P.: A thermal theory for unifying and designing transparency, concentrating and cloaking. J. Appl. Phys. {\bf 123}, 115117 (2018)

\bibitem{Yfb-YangESEE19} Yang, F.B., Xu, L.J., Huang, J.P.: Thermal illusion of porous media with convection-diffusion process: Transparency, concentrating, and cloaking. ES Energy Environ. {\bf 6}, 45-50 (2019)  

\bibitem{Yfb-ZhuangSCPMA22} Zhuang, P.F., Xu, L.J., Tan, P., Ouyang, X.P., Huang, J.P.: Breaking efficiency limit of thermal concentrators by conductivity couplings. Sci. China Phys. Mech. Astron. {\bf 65}, 117007 (2022)




\bibitem{Yfb-JinIJHMT20} Jin, P., Xu, L.J., Jiang, T., Zhang, L., Huang, J.P.: Making thermal sensors accurate and invisible with an anisotropic monolayer scheme. Int. J. Heat Mass Transf. {\bf 163}, 120437 (2020)

\bibitem{Yfb-XuEPL20b} Xu, L.J., Huang, J.P., Jiang, T., Zhang, L., Huang, J.P.: Thermally invisible sensors. Europhys. Lett. {\bf 132}, 14002 (2020)

\bibitem{Yfb-JinIJHMT21} Jin, P., Yang, S., Xu, L.J., Dai, G.L., Huang, J.P., Ouyang, X.P.: Particle swarm optimization for realizing bilayer thermal sensors with bulk isotropic materials. Int. J. Heat Mass Transfer {\bf 172}, 121177 (2021)

\bibitem{Yfb-WangEPL21} Wang, C.Q., Xu, L.J., Jiang, T., Zhang, L., Huang, J.P.: Multithermally invisible cloaks and sensors with complex shapes. Europhys. Lett. {\bf 133}, 20009 (2021)


\bibitem{Yfb-YangPRAP20} Yang, F.B., Tian, B.Y., Xu, L.J., Huang, J.P.: Experimental demonstration of thermal chameleonlike rotators with transformation-invariant metamaterials. Phys. Rev. Appl. {\bf 14}, 054024 (2020)

\bibitem{Yfb-QiuCTP14} Gao, Y., Wang, W.H., Huang, J.P.: Transformation electrics: Cloaking and rotating electric current. Commun. Theor. Phys. {\bf 61}, 517-520 (2014)

\bibitem{Yfb-DaiJAP18} Dai, G.L., Huang, J.P.: A transient regime for transforming thermal convection: Cloaking, concentrating and rotating creeping flow and heat flux. J. Appl. Phys. {\bf 124}, 235103 (2018)

\bibitem{Yfb-HuangCP04} Huang, J.P., Yu, K.W.: Many-body dipole-induced dipole model for electrorheological fluids. Chin. Phys. {\bf 13}, 1065-1069 (2004)

\bibitem{Yfb-HuangPRE04d} Huang, J.P.: Statistical-mechanical theory of the overall magnetic properties of mesocrystals. Phys. Rev. E {\bf 70}, 041403 (2004)

\bibitem{Yfb-HuangPRE04b} Huang, J.P., Yu, K.W.: Interparticle force in electrorheological solids: Many-body dipole-induced dipole model. Phys. Rev. E {\bf 70}, 061401 (2004)

\bibitem{Yfb-HuangPRE05a} Huang, J.P., Yu, K.W., Gu, G.Q., Karttunen, M., Dong, L.: Reply to Comment on the use of the method of images for calculating electromagnetic responses of interacting spheres. Phys. Rev. E {\bf 72}, 023402 (2005)

\bibitem{Yfb-XuPLA06} Xu, B., Huang, J.P., Yu, K.W.: Theory of nonlinear ac responses of inhomogeneous two-component composite films. Phys. Lett. A {\bf 357}, 475-478 (2006)

\bibitem{Yfb-ChenJPA07} Chen, Y.W., Zhang, L.F., Huang, J.P.: The Watts-Strogatz network model developed by including degree distribution: Theory and computer simulation. J. Phys. A Math. Theor. {\bf 40}, 8237-8246 (2007)

\bibitem{Yfb-WangPNAS09} Wang, W., Chen, Y., Huang, J.P.: Heterogeneous preferences, decision-Making capacity and phase transitions in a complex adaptive system. Proc. Natl. Acad. Sci. U.S.A. {\bf 106}, 8423-8428 (2009)

\bibitem{Yfb-WangOL10} Wang, G., Huang, J.P., Yu, K.W.: Nontrivial Bloch oscillations in waveguide arrays with second-order coupling. Opt. Lett. {\bf 35}, 1908-1910 (2010)

\bibitem{Yfb-ZhaoPNAS11} Zhao, L., Yang, G., Wang, W., Chen, Y., Huang, J.P., Ohashi, H., Stanley, H. E.: Herd behavior in a complex adaptive system. Proc. Natl. Acad. Sci. U.S.A. {\bf 108}, 15058-15063 (2011)

\bibitem{Yfb-SongARCS12} Song, K.Y., An, K.N., Yang, G., Huang, J.P.: Risk-return relationship in a complex adaptive system. PLoS One {\bf 7}, e33588 (2012)

\bibitem{Yfb-liang2013pre} Liang, Y., An, K.N., Yang, G., Huang, J.P.: Contrarian behavior in a complex adaptive system. Phys. Rev. E {\bf 87}, 012809 (2013)

\bibitem{Yfb-Liang2013fopw} Liang, Y., Huang, J.P.: Robustness of critical points in a complex adaptive system: Effects of hedge behavior. Front. Phys. {\bf 8}, 461-466 (2013)

\bibitem{Yfb-QiuFP14} Zheng, W.Z., Liang, Y., Huang, J.P.: Equilibrium state and non-equilibrium steady state in an isolated human system. Front. Phys. {\bf 9}, 128-135 (2014)

\bibitem{Yfb-QiuPLA14} Liang, Y., An, K.N., Yang, G., Huang, J.P.: A possible human counterpart of the principle of increasing entropy. Phys. Lett. A {\bf 378}, 488-493 (2014)

\bibitem{Yfb-QiuCPB14} Liang, Y., Huang, J.P.: Statistical physics of human beings in games: Controlled experiments. Chin. Phys. B {\bf 23}, 078902 (2014)

\bibitem{Yfb-Tan12} Li, X.H., Yang, G., Huang, J.P.: Chaotic-periodic transition in a two-sided minority game. Front. Phys. {\bf 11}, 118901 (2016)

\bibitem{Yfb-Tan5} Li, X.H., Yang, G., An, K.N., Huang, J.P.: Human behavioral regularity, fractional Brownian motion, and exotic phase transition. Phys. Lett. A {\bf 380}, 2912-2919 (2016)

\bibitem{Yfb-Tan2} Xin, C., Zhang, H.S., Huang, J.P.: Complex network approach to classifying classical piano compositions. EPL {\bf 116}, 18008 (2016)

\bibitem{Yfb-XinPA17} Xin, C., Yang, G., Huang, J.P.: Ising game: Nonequilibrium steady states of resource-allocation systems. Physica A {\bf 471}, 666-673 (2017)

\bibitem{Yfb-XinFP17} Xin, C., Huang, J.P.: Recent progress in econophysics: Chaos, leverage, and business cycles as revealed by agent-based modeling and human experiments. Front. Phys. {\bf 12}, 128910 (2017)

\bibitem{Yfb-JiPA18} Ji, Q., Xin, C., Tang, S.X., Huang, J.P.: Symmetry associated with symmetry break: Revisiting ants and humans escaping from multiple-exit rooms. Physica A {\bf 492}, 941-947 (2018)







\bibitem{Yfb-HuangJMMM05} Huang, J.P., Wang, Z.W., Holm, C.: Structure and magnetic properties of mono-and bi-dispersed ferrofluids as revealed by simulations. J. Magn. Magn. Mater. {\bf 289}, 234-237 (2005)

\bibitem{Yfb-HuangSSC00} Huang, J.P., Gao, L., Li, Z.Y.: Temperature effect on nonlinear optical response in metal/dielectric composite with interfacial layer. Solid State Commun. {\bf 115}, 347-352 (2000)

\bibitem{Yfb-HuangCTP01-2} Huang, J.P., Gao, L., Li, Z.Y.: Optical response of metal/dielectric composite containing interfacial layers. Commun. Theor. Phys. {\bf 36}, 251-256 (2001)

\bibitem{Yfb-HuangCTP01-1} Huang, J.P., Li, Z.Y.: Effects of particle shape and microstructure on effective nonlinear response. Commun. Theor. Phys. {\bf 36}, 365-369 (2001)

\bibitem{Yfb-PanPB01} Pan, T., Huang, J.P., Li, Z.Y.: Optical bistability in metal-dielectric composite with interfacial layer. Physica B {\bf 301}, 190-195 (2001)

\bibitem{Yfb-HuangPRE01} Huang, J.P., Wan, J.T.K., Lo, C.K., Yu, K.W.: Nonlinear ac response of anisotropic composites. Phys. Rev. E {\bf 64}, 061505 (2001)

\bibitem{Yfb-HuangJPCM02} Huang, J.P., Yu, K.W.: First-principles approach to electrorotation assay. J. Phys. Condens. Matter {\bf 14}, 1213-1221 (2002)

\bibitem{Yfb-HuangPRE02} Huang, J.P., Yu, K.W., Gu, G.Q.: Electrorotation of a pair of spherical particles. Phys. Rev. E {\bf 65}, 021401 (2002)

\bibitem{Yfb-HuangCTP02} Huang, J.P., Yu, K.W., Lei, J., Sun, H.: Spectral representation theory for dielectric behavior of nonspherical cell suspensions. Commun. Theor. Phys. {\bf 38}, 113-120 (2002)

\bibitem{Yfb-HuangPLA02} Huang, J.P., Yu, K.W., Gu, G.Q.: Electrorotation of colloidal suspensions. Phys. Lett. A {\bf 300}, 385-391 (2002)

\bibitem{Yfb-HuangCTP03} Huang, J.P., Yu, K.W.: Dielectric behavior of oblate spheroidal particles: Application to erythrocytes suspensions. Commun. Theor. Phys. {\bf 39}, 506-512 (2003)

\bibitem{Yfb-HuangPRE03-2} Huang, J.P., Karttunen, M., Yu, K.W., Dong, L.: Dielectrophoresis of charged colloidal suspensions. Phys. Rev. E {\bf 67}, 021403 (2003)

\bibitem{Yfb-GaoPRE03} Gao, L., Huang, J.P., Yu, K.W.: Theory of ac electrokinetic behavior of spheroidal cell suspensions with an intrinsic dispersion. Phys. Rev. E {\bf 67}, 021910 (2003)

\bibitem{Yfb-HuangJAP03} Huang, J.P., Gao, L., Yu, K.W.: Nonlinear alternating current response of colloidal suspension with an intrinsic dispersion. J. Appl. Phys. {\bf 93}, 2871-2875 (2003)

\bibitem{Yfb-HuangPRE03-1} Huang, J.P., Yu, K.W., Gu, G.Q., Karttunen, M.: Electrorotation in graded colloidal suspensions. Phys. Rev. E {\bf 67}, 051405 (2003)

\bibitem{Yfb-GaoEPJB03} Gao, L., Huang, J.P., Yu, K.W.: Giant enhancement of optical nonlinearity in mixtures of graded particles with dielectric anisotropy. Eur. Phys. J. B {\bf 36}, 475-484 (2003)

\bibitem{Yfb-DongJAP04} Dong, L., Huang, J.P., Yu, K.W., Gu, G.Q.: Dielectric response of graded spherical particles of anisotropic materials. J. Appl. Phys. {\bf 95}, 621-624 (2004)

\bibitem{Yfb-KoJPCM04} Ko, Y.T.C., Huang, J.P., Yu, K.W.: Dielectric behaviour of single-shell spherical cells with an intrinsic dispersion. J. Phys. Condens. Matter {\bf 16}, 499-509 (2004)

\bibitem{Yfb-GaoPRB04} Gao, L. Huang, J.P., Yu, K.W.: Effective nonlinear optical properties of composite media of graded spherical particles. Phys. Rev. B {\bf 69}, 075105 (2004)

\bibitem{Yfb-HuangPRE04g} Huang, J.P., Gao, L., Yu, K.W., Gu, G.Q.: Nonlinear alternating current responses of graded materials. Phys. Rev. E {\bf 69}, 036605 (2004)

\bibitem{Yfb-LiuPLA04} Liu, R.M., Huang, J.P.: Theory of the dielectrophoretic behavior of clustered colloidal particles in two dimensions. Phys. Lett. A {\bf 324}, 458-464 (2004)

\bibitem{Yfb-DongJAP04-2} Dong, L., Huang, J.P., Yu, K.W.: Theory of dielectrophoresis in colloidal suspensions. J. Appl. Phys. {\bf 95}, 8321-8326 (2004)

\bibitem{Yfb-HuangCPL04} Huang, J.P.: Force acting on the microparticles in electrorheological solids under the application of a nonuniform ac electric field. Chem. Phys. Lett. {\bf 390} 380-383 (2004)

\bibitem{Yfb-HuangPRE04f} Huang, J.P., Karttunen, M., Yu, K.W., Dong, L., Gu, G.Q.: Electrokinetic behavior of two touching inhomogeneous biological cells and colloidal particles: Effects of multipolar interactions. Phys. Rev. E {\bf 69}, 051402 (2004)

\bibitem{Yfb-HuangJPCM04} Huang, J.P.: New mechanism for harmonic generation in magnetorheological fluids. J. Phys. Condes. Matter {\bf 16}, 7889-7894 (2004)

\bibitem{Yfb-HuangPLA04} Huang, J.P., Yu, K.W.: AC electrokinetics of the microparticles in electrorheological solids. Phys. Lett. A {\bf 333}, 347-353 (2004)

\bibitem{Yfb-KoEPJE04} Ko, Y.T.C., Huang, J.P., Yu, K.W.: Dielectric behaviour of graded spherical cells with an intrinsic dispersion. Eur. Phys. J. E {\bf 14}, 97-104 (2004)

\bibitem{Yfb-HuangAPL04} Huang, J.P., Yu, K.W.: Optical nonlinearity enhancement of graded metallic films. Appl. Phys. Lett. {\bf 85}, 94-96 (2004)

\bibitem{Yfb-HuangPRE04e} Huang, J.P., Yu, K.W., Karttunen, M.: Nonlinear alternating current responses of dipolar fluids. Phys. Rev. E {\bf 70}, 011403 (2004)   

\bibitem{Yfb-HuangJCP04} Huang, J.P., Yu, K.W.: Nonlinear ac responses of electro-magnetorheological fluids. J. Chem. Phys. {\bf 121}, 7526-7532 (2004)



\bibitem{Yfb-HuangPRE04c} Huang, J.P.: New nonlinear dielectric materials: Linear electrorheological fluids under the influence of electrostriction. Phys. Rev. E {\bf 70}, 042501 (2004)


\bibitem{Yfb-HuangJPCB04} Huang, J.P.: Theory of electrical conductivities of ferrogels. J. Phys. Chem. B {\bf 108}, 13901-13904 (2004)

\bibitem{Yfb-WangJAP03} Wang, G., Tian, W.J., Huang, J.P.: Response of ferrogels subjected to an AC magnetic field. J. Phys. Chem. B {\bf 110}, 10738 (2006)

\bibitem{Yfb-ShenCPL06} Shen, M., Cao, J.G., Xue, H.T., Huang, J.P., Zhou, L.W.: Structure of polydisperse electrorheological fluids: Experiment and theory. Chem. Phys. Lett. {\bf 423}, 165-169 (2006)

\bibitem{Yfb-WangCPL06} Wang, G., Huang, J.P.: Nonlinear magnetic susceptibility of colloidal ferrofluids. Chem. Phys. Lett. {\bf 421}, 544-548 (2006)

\bibitem{Yfb-HuangAPL06} Huang, J.P.: Ferrogels under the influence of a nonuniform magnetic field. Appl. Phys. Lett. {\bf 88}, 091912 (2006)

\bibitem{Yfb-HuangJPD06} Huang, J.P.: The behavior of a single magnetic particle under the influence of an AC magnetic field. J. Phys. D: Appl. Phys. {\bf 39}, 2455-2460 (2006)

\bibitem{Yfb-WangPRE06} Wang, G., Huang, J.P.: The behavior of interacting magnetic nanoparticles subjected to an AC magnetic field. Phys. Rev. E {\bf 74}, 061406 (2006)

\bibitem{Yfb-ZhangPRE06} Zhang, G.J., Yu, K.W., Huang, J.P.: Electric double layers in electrolyte solutions under shear flow: Effects of shear rate. Phys. Rev. E {\bf 74}, 031501 (2006)

\bibitem{Yfb-ZhangCPL07} Zhang, G.J., Yu, K.W., Huang, J.P.: Nonlinear electrokinetic responses of an electrolyte solution subjected to an AC electric field. Chem. Phys. Lett. {\bf 448}, 1-5 (2007)

\bibitem{Yfb-ZhangPRE07} Zhang, G.J., Yu, K.W., Huang, J.P.: AC electrokinetic behavior of an electrolyte solution in a microchannel. Phys. Rev. E {\bf 76}, 056304 (2007)



\bibitem{Yfb-SamalJCP17} Samal, S.: Thermal plasma technology: The prospective future in material processing. J. Clean Prod. {\bf 142}, 3131-3150 (2017)

\bibitem{Yfb-LiangAEM18} Liang, H., Ming, F. Alshareef, H.N.: Applications of Plasma in Energy Conversion and Storage Materials. Adv. Energy Mater. {\bf 8}, 1801804 (2018)

\bibitem{Yfb-TamIEEE20} Tamura, H., Tetsuka, T., Kuwahara, D., Shinohara, S.: Study on uniform plasma generation mechanism of electron cyclotron resonance etching reactor. IEEE T. Plasma Sci. {\bf 48}, 3606-3615 (2020)

\bibitem{Yfb-LiAST21} Li, M., Wang, Z., Xu, R., Zhang, X., Chen, Z., Wang, Q.: Advances in plasma-assisted ignition and combustion for combustors of aerospace engines. Aerosp. Sci. Technol. {\bf 117}, 106952 (2021)




\bibitem{Yfb-HuAM18} Hu, R., Zhou, S., Li, Y., Lei ,D.Y., Luo, X., Qiu, C.W.: Illusion thermotics. Adv. Mater. {\bf 30}, 1707237 (2018)

\bibitem{Yfb-HuAM19} Hu, R., Huang, S., Wang, M., Luo, X., Shiomi, J., Qiu, C.W.: Encrypted thermal printing with regionalization transformation. Adv. Mater. {\bf 31}, 1807849 (2019)

\bibitem{Yfb-ZhangCPL21} Zhang, J., Huang, S., Hu, R.: Adaptive radiative thermal camouflage via synchronous heat conduction. Chin. Phys. Lett. {\bf 38}, 010502 (2021)

\bibitem{Yfb-ZhangCPL22} Zhang, Z.R., Huang, J.P.: Transformation plasma physics. Chin. Phys. Lett. {\bf 39}, 075201 (2022)

\bibitem{Yfb-CuiJAP19} Cui, S., Wu, Z., Lin, H., Xiao, S., Zheng, B., Liu, L., An, X., Fu, R.K.Y., Tian, X., Tan, W., Chu, P.K.: Hollow cathode effect modified time-dependent global model and high-power impulse magnetron sputtering discharge and transport in cylindrical cathode. J. Appl. Phys. {\bf 125}, 063302 (2019)

\bibitem{Yfb-DaiFP21} Dai, G.L.: Designing nonlinear thermal devices and metamaterials under the Fourier law: A route to nonlinear thermotics. Front. Phys. {\bf 16}, 53301 (2021)


\bibitem{Yfb-RodPRAP21} Rodríguez, J.A., Abdalla, A.I., Wang, B., Lou, B., Fan, S., Cappelli, M.A.: Inverse design of plasma metamaterial devices for optical computing. Phys. Rev. Appl. {\bf 16}, 014023 (2021)

\bibitem{Yfb-InaJAP21} Inami, C., Kabe, Y., Noyori, Y., Iwai, A., Bambina, A., Miyagi, S., Sakai, O.: Experimental observation of multi-functional plasma-metamaterial composite for manipulation of electromagnetic-wave propagation. J. Appl. Phys. {\bf 130}, 043301 (2021)

\bibitem{Yfb-ZhouCS21} Zhou, X., Xu, G., Zhang, H.: Binary masses manipulation with composite bilayer metamaterial. Compos. Struct. {\bf 267}, 113866 (2021)

\bibitem{Yfb-ResSR16} Restrepo-Flórez, J.M., Maldovan, M.: Mass separation by metamaterials. Sci. Rep. {\bf 6}, 21971 (2016)

\bibitem{Yfb-HuPRX20} Hu, R., Iwamoto, S., Feng, L., Ju, S., Hu, S., Ohnishi, M., Nagai, N., Hirakawa, K., Shiomi, J.: Machine-learning-optimized aperiodic superlattice minimizes coherent phonon heat conduction. Phys. Rev. X {\bf 267}, 021050 (2020)


\bibitem{Yfb-NaraAM12} Narayana, S., Sato, Y.: DC magnetic cloak. Adv. Mater. {\bf 24}, 71-74 (2012)


\bibitem{Yfb-LanSR15} Lan, C., Yang, Y., Geng, Z., Li, B., Zhou, J.: Electrostatic field invisibility cloak. Sci. Rep. {\bf 5}, 16416 (2015)


\bibitem{Yfb-HuangIEEE15} Huang, C.W., Chen, Y.C., Nishimura, Y.: Particle-in-cell simulation of plasma sheath dynamics with kinetic ions. IEEE T. Plasma Sci. {\bf 43}, 675-682 (2015)


\bibitem{Yfb-YuFP15} Yu, Z.Z., Xiong, G.H., Zhang, L.F.: A brief review of thermal transport in mesoscopic systems from nonequilibrium Green’s function approach. Front. Phys. {\bf 16}, 43201 (2015)


\bibitem{Yfb-XingCPL21} Xing, G., Zhao, W., Hu, R., Luo, X.: Spatiotemporal modulation of thermal emission from thermal-hysteresis vanadium dioxide for multiplexing thermotronics functionalities. Chin. Phys. Lett. {\bf 38}, 124401 (2021)


\bibitem{Yfb-Torrent2018} Torrent, D., Poncelet, O., Batsale, J.-C.: Nonreciprocal thermal material by spatiotemporal modulation. Phys. Rev. Lett. {\bf 120}, 125501 (2018)

\bibitem{Yfb-Zang2019} Zang, J.W., Correas-Serrano, D., Do, J.T.S., Liu, X., Alvarez-Melcon, A., Gomez-Diaz, J.S.: Nonreciprocal wavefront engineering with time-modulated gradient metasurfaces. Phys. Rev. Appl. {\bf 11}, 054054 (2019)

\bibitem{Yfb-Guo2019} Guo, X.X., Ding, Y.M., Duan, Y., Ni, X.J.: Nonreciprocal metasurface with space-time phase modulation. Light-Sci. Appl. {\bf 8}, 123 (2019)



\bibitem{Yfb-Camacho2020a} Camacho, M., Edwards, B., Engheta, N.: Achieving asymmetry and trapping in diffusion with spatiotemporal metamaterials. Nat. Commun. {\bf 11}, 3733 (2020)

\bibitem{Yfb-Li2022} Li, J., Li, Y., Cao, P.-C., Qi, M., Zheng, X., Peng, Y.-G., Li, B., Zhu, X.-F., Alù, A., Chen, H., Qiu, C.-W.: Thermal reciprocity in metamaterials. Nat. Commun. {\bf 13}, 167 (2022)


\bibitem{Yfb-GongPRX18} Gong, Z.P., Ashida, Y., Kawabata, K., Takasan, K., Higashikawa, S., Ueda, M.: Topological phases of non-Hermitian systems. Phys. Rev. X {\bf 8}, 031079 (2018)

\bibitem{Yfb-KawabataPRX19} Kawabata, K., Shiozaki, K., Ueda, M., Sato, M.: Symmetry and topology in non-Hermitian physics. Phys. Rev. X {\bf 9}, 041015 (2019)

\bibitem{Yfb-KlitzingPRL80} Klitzing, K.v., Dorda, G., Pepper, M.: New method for high-accuracy determination of the fine-structure constant based on quantized Hall resistance. Phys. Rev. Lett. {\bf 45}, 494 (1980)

\bibitem{Yfb-Klitzing05} Klitzing, K.v.: Developments in the quantum Hall effect. Philos. Trans. R. Soc. London, Ser. A {\bf 363}, 2203 (2005)

\bibitem{Yfb-HasanRMP10} Hasan, M.Z., Kane, C.L.: Colloquium: Topological insulators. Rev. Mod. Phys. {\bf 82}, 3045 (2010)

\bibitem{Yfb-QiRMP11} Qi, X.-L., Zhang, S.-C.: Topological insulators and superconductors. Rev. Mod. Phys. {\bf 83}, 1057 (2011)

\bibitem{Yfb-LuNP14} Lu, L., Joannopoulos, J.D., Solja${\rm{\breve{c}}}$i${\rm{\acute{c}}}$, M.: Topological photonics. Nat. Photonics {\bf 8}, 821 (2014)

\bibitem{Yfb-OzawaRMP19} Ozawa, T., Price, H.M., Amo, A., Goldman, N., Hafezi, M., Lu, L., Rechtsman, M.C., Schuster, D., Simon, J., Zilberberg, O., Carusotto, I.: Topological photonics. Rev. Mod. Phys. {\bf 91}, 015006 (2019)

\bibitem{Yfb-YangPRL15} Yang, Z.J., Gao, F., Shi, X.H., Lin, X., Gao, Z., Chong, Y.D., Zhang, B.L.: Topological acoustics. Phys. Rev. Lett. {\bf 114}, 114301 (2015)

\bibitem{Yfb-XueNRM22} Xue, H.R., Yang, Y.H., Zhang, B.L.: Topological acoustics. Nat. Rev. Mater. {\bf 7}, 974 (2022)

\bibitem{Yfb-HuberNP16} Huber, S.D.: Topological mechanics. Nat. Phys. {\bf 12}, 621 (2016)



\bibitem{Yfb-ParkerJPP21} Parker, J.B.: Topological phase in plasma physics. J. Plasma Phys. {\bf 87}, 835870202 (2021)









\end{thebibliography}
\end{document}